\documentclass[10pt,twocolumn,letterpaper]{article}

\usepackage{iccv}              
\usepackage[accsupp]{axessibility} 

\usepackage[dvipsnames]{xcolor}

\usepackage{amsmath,amssymb}
\usepackage{amssymb} 
\usepackage{pifont}  
\usepackage{booktabs}
\usepackage{multirow}
\usepackage{multicol}
\usepackage[mathscr]{euscript}
\usepackage[dvipsnames]{xcolor}
\usepackage{colortbl}
\usepackage{makecell}
\usepackage{subcaption}
\usepackage{graphicx}

\newcommand{\Checkmark}{\ding{51}}   
\newcommand{\XSolidBrush}{\ding{55}}  

\newcommand{\websitelink}{https://anonymous.4open.science/w/SoulDance-BBD3/}

\definecolor{iccvblue}{rgb}{0.21,0.49,0.74}
\usepackage[pagebackref,breaklinks,colorlinks,allcolors=iccvblue]{hyperref}

\def\paperID{7644} 
\def\confName{ICCV}
\def\confYear{2025}

\title{Music-Aligned Holistic 3D Dance Generation via Hierarchical Motion Modeling}

\author{Xiaojie Li$^{1, 2*}$, Ronghui Li$^1$, Shukai Fang$^2$, Shuzhao Xie$^1$, Xiaoyang Guo$^2$, \\ Jiaqing Zhou$^2$, Junkun Peng$^1$, Zhi Wang$^{1\dagger}$\\
    $^1$Shenzhen International Graduate School, Tsinghua University  
    $^2$ByteDance Games\\
		\small{\{li-xj23, lrh22, xsz24, pjk20\}@mails.tsinghua.edu.cn, wangzhi@sz.tsinghua.edu.cn, \{fangshukai, beichuan, jiashu\}@bytedance.com}
        }
\begin{document}
\twocolumn[{
\renewcommand\twocolumn[1][]{#1}
\maketitle
\begin{center}
    \centering
    \captionsetup{type=figure}
    \includegraphics[width=1\linewidth]{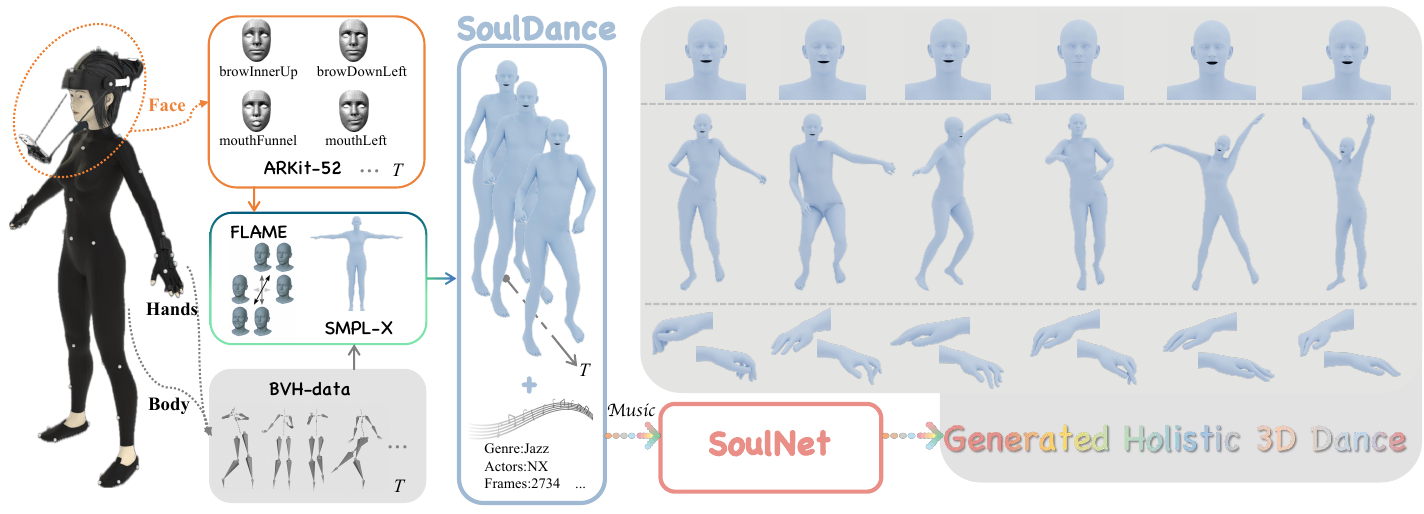}

    \captionof{figure}{We introduce \textbf{\textit{SoulDance}}, a high-quality, comprehensive dance dataset that incorporates body motions, hand gestures, and facial expressions. The dataset processing pipeline (left) consists of separate steps for capturing facial expressions and body-hand movements. Moreover, we present \textbf{\textit{SoulNet}}, the first framework able to generate expressive and holistic dance, as demonstrated in the results (right).}
    \label{fig:teaser_image}
\end{center}
}]

\newcommand{\lrh}[1]{{\color{blue}[\bf \em lrh: {#1}]}}
\newcommand{\fsk}[1]{{\color{cyan}{#1}}}

\def\thefootnote{*}\footnotetext{This work was partially conducted during an internship at ByteDance.}
\def\thefootnote{$\dagger$}\footnotetext{Corresponding author.}

\begin{abstract}  
Well-coordinated, music-aligned holistic dance enhances emotional expressiveness and audience engagement. 
However, generating such dances remains challenging due to the scarcity of holistic 3D dance datasets, the difficulty of achieving cross-modal alignment between music and dance, and the complexity of modeling interdependent motion across the body, hands, and face.
To address these challenges, we introduce SoulDance, a high-precision music-dance paired dataset captured via professional motion capture systems, featuring meticulously annotated holistic dance movements.  
Building on this dataset, we propose SoulNet, a framework designed to generate music-aligned, kinematically coordinated holistic dance sequences. SoulNet consists of three principal components: (1) Hierarchical Residual Vector Quantization, which models complex, fine-grained motion dependencies across the body, hands, and face; (2) Music-Aligned Generative Model, which composes these hierarchical motion units into expressive and coordinated holistic dance; (3) Music-Motion Retrieval Module, a pre-trained cross-modal model that functions as a music-dance alignment prior, ensuring temporal synchronization and semantic coherence between generated dance and input music throughout the generation process. 
Extensive experiments demonstrate that SoulNet significantly surpasses existing approaches in generating high-quality, music-coordinated, and well-aligned holistic 3D dance sequences. Additional resources are available at: \url{https://xjli360.github.io/SoulDance}.
\end{abstract}

\section{Introduction}
\label{sec:intro}

Dance represents a fundamental form of artistic expression and cultural heritage across human civilizations. As a universal language transcending verbal communication, dance integrates complex body movements, hand gestures, and facial expressions to convey emotions and narratives~\cite{fink2021evolution, kico2018digitization}. With the rapid advancement of digital entertainment, including video games, virtual reality, and digital performances, the demand for realistic and expressive 3D dance animation has grown substantially. Conventional methodologies for acquiring dance assets rely heavily on professional dancers and sophisticated motion capture systems, making the process labor intensive, time-consuming, and prohibitively expensive. This practical constraint has motivated significant research interest in dance generation.

Early methods utilizing motion graphs~\cite{arikan2002interactive, kim2003rhythmic, shiratori2006dancing, kim2006making, tang2018dance} ensure local motion quality but fail to capture intrinsic music-dance relationships. Learning-based approaches like FACT~\cite{li2021ai} and EDGE~\cite{tseng2023edge} enhance local dance quality through window-based learning but neglect global choreographic coherence. More recent methods such as Bailando~\cite{siyao2022bailando} employ VQ-VAE~\cite{van2017neural} with reinforcement learning for rhythm alignment, but suffer from training complexity and limited generalizability. While FineNet~\cite{li2023finedance} advances by modeling both body and hand motions, it lacks the capability to generate corresponding facial expressions, resulting in emotionally incomplete performances.

Nonetheless, despite recent advances, the generation of well-coordinated and holistic dance sequences that are synchronized with music remains a challenging task, primarily due to three fundamental limitations. The first constraint refers to dataset inadequacies in which existing music-dance datasets either lack crucial components of dance performance or suffer from quality issues. While datasets like AIST++\cite{li2021ai} provide body-only movements with musical correspondence, and FineDance\cite{li2023finedance} and Choreomaster~\cite{chen2021choreomaster} incorporate fine-grained finger movements, they omit facial expressions—a vital element for conveying emotion in dance. Furthermore, most datasets rely on pose estimation from videos, introducing inherent inaccuracies that compromise data quality. The second limitation concerns modeling constraints where current approaches fail to generate coordinated dance movements due to insufficient capacity for capturing complex hierarchical interdependencies between body, hands and face. The third limitation lies in the lack of cross-modal alignment between music and dance, which results in poor synchronization between the two modalities.


To overcome these challenges, we present a comprehensive solution encompassing both dataset and methodological innovations. We introduce \textbf{\textit{SoulDance}}, a high-quality dataset comprising 12.5 hours of paired music and dance data captured using professional motion capture systems equipped with 15 optical cameras, body motion capture suits, data gloves, and a facial expression tracking system. Unlike existing datasets, SoulDance is a holistic dance dataset that includes body movements, hand gestures, and facial expressions, captured from professional dancers across 284 music segments spanning 15 genres. The data is represented in multiple formats, including skeleton-level BVH files, SMPL-X~\cite{SMPL-X:2019} body models, and FLAME~\cite{FLAME:SiggraphAsia2017} face parameters, providing unprecedented quality and diversity for dance research.

Building on this dataset, we propose a holistic dance generation framework, \textbf{\textit{SoulNet}}, comprising three principal components that address coordinated movement and cross-modal alignment challenges. First, to resolve coordination issues, we introduce \textit{Hierarchical Residual Vector Quantization}, which extends RVQ techniques~\cite{barnes1996advances,huijben2024residual} to model complex spatial relationships among body movements, hand gestures, and facial expressions through a multi-layer encoding structure and body-hand-face chain design. Second, to address alignment challenges, we develop a \textit{Music-Aligned Generative Model}—a transformer-based architecture that composes hierarchical dance units into coherent sequences while maintaining musical synchronization. Third, We enhance this with a pre-trained \textit{Music-Motion Retrieval} module that leverages extensive public music-dance datasets to provide alignment priors guiding the generation process. Additionally, to rigorously evaluate our approach, we introduce two novel evaluation metrics: the EmotionAlign Score, which quantifies alignment between facial expressions and musical emotional tone, and the MMR-Matching Score, which enables fine-grained assessment of synchronization between musical rhythm and dance movements. These metrics offer more precise and comprehensive evaluation compared to existing measures such as the Genre-Matching Score~\cite{li2023finedance}, which primarily relies on coarse genre labels.

Overall, our main contributions are as follows:
\begin{itemize}
    \item We present \textit{SoulDance}, a new high-quality, music-paired, holistic dance dataset, the largest to date, encompassing body movements, hand gestures, and facial expressions.
    \item We propose \textit{SoulNet}, a framework that leverages learned music-dance alignment priors to achieve coordinated holistic dance modeling with rhythmic and emotional consistency between dance and music.
    \item We introduce two new metrics, the EmotionAlign Score and the MMR-Matching Score, to provide a more precise evaluation of facial and body motion alignment with music. Our extensive experiments on multiple datasets demonstrate the outstanding performance of our method on both these metrics and public benchmarks.
\end{itemize}
\section{Related Work}
\label{sec:related_work}

\noindent\textbf{3D Dance Datasets.} Many existing dance datasets are constrained by either quantity or quality. 
Some datasets reconstruct 3D poses from multi-view 2D videos, often compromising pose accuracy~\cite{li2021ai, le2023music}. Motion capture datasets~\cite{valle2021transflower, tang2018dance, zhuang2022music2dance} featuring professional dancers typically achieve high quality but are limited in duration due to high production costs. These datasets often lack detailed hand movements and exhibit insufficient diversity. FineDance~\cite{li2021ai} offers a 14.6-hour motion capture dataset with body and hand motions, but it does not include facial expressions. In contrast, our proposed SoulDance provides comprehensive holistic motion data, including body, hands, and facial expressions, with 12.5 hours of synchronized music and dance data. \textit{SoulDance} significantly enhances both the quality and diversity of the data.


\noindent\textbf{Motion Quantization.} 
Vector Quantization (VQ) ~\cite{gray1998quantization} has been applied in generative modeling via the Vector Quantized Variational Autoencoder (VQ-VAE)~\cite{van2017neural}, where it enables discrete latent by selecting specific codebook entries for encoding. This approach inspired work in motion generation~\cite{guo2022tm2t, siyao2022bailando}, which employs VQ-VAE to encode human motion as discrete tokens. To further improve motion representation, T2M-GPT~\cite{zhang2023t2m} introduced EMA and code reset techniques to improve the performance of VQ-based models. HumanTomato~\cite{lu2023humantomato} and DanceMeld~\cite{gao2023dancemeld} employ hierarchical VQ to capture more detailed motion features. Despite these advances, the inherent error introduced in the quantization process remains a challenge. To mitigate this, MoMask~\cite{guo2024momask} adapted Residual Vector Quantization (RVQ)~\cite{zeghidour2021soundstream, borsos2023audiolm}, which iteratively quantizes both the primary vector and its residuals, thus refining the approximation in each step and reducing errors. However, these methods often lack the ability to capture hierarchical relationships crucial for modeling complex holistic interactions. To address this limitation, we propose Hierarchical Residual Vector Quantization, an extension of RVQ that incorporates a structured approach for holistic modeling, improving the fidelity of holistic motion generation.

\noindent\textbf{Dance Generation.} 
Early synthesis-based approaches primarily focused on retrieving dance motions based on music cues~\cite{arikan2002interactive, kim2003rhythmic, shiratori2006dancing, kim2006making, tang2018dance, au2022choreograph, chen2021choreomaster}. However, these methods struggled to capture the complex relationships between music and dance movements, resulting in a limited expressive range. Recent advancements in generative models, such as GAN-based and transformer-based frameworks, have improved the quality of generated dance~\cite{li2021ai, li2022danceformer, siyao2022bailando, zhang2024bidirectional}. Despite these improvements, these models often produce unrealistic and homogeneous choreography, lacking the dynamism of authentic dance movements. Diffusion-based models have emerged with strong generative capabilities, enabling the creation of more vivid dance sequences~\cite{li2022danceformer, tseng2023edge, li2023finedance, li2024lodge}. However, directly mapping music to high-dimensional joints sequences can introduce instability, resulting in non-standard poses outside the dance manifold. Furthermore, these models often emphasize body or hand motions while neglecting facial expressions, leading to unnatural and emotionally detached dance sequences. Our proposed \textit{SoulNet} addresses these challenges by capturing spatial correlations across holistic components and ensuring precise dance-music alignment, thereby enhancing the authenticity and emotional depth of the generated dance.

\begin{table*}[ht]
    \setlength\tabcolsep{3pt}
    \centering
    \resizebox{\textwidth}{!}{
        \begin{tabular}{lccccccccccc}
            \toprule [1pt] \noalign{\smallskip}
            Dataset &  Pos/Rot & Joints num & Hand & Face & Genres & Mocap& Single Dance   & Group Dance & SMPL & \makecell[c]{Total\\hours} & \makecell[c]{avg Sec\\per Seq}\\
            \hline\noalign{\smallskip}\noalign{\smallskip}
            PMSD~\cite{valle2021transflower} & \Checkmark/\XSolidBrush & 52 & \XSolidBrush & \XSolidBrush &14 & \Checkmark & \Checkmark & \XSolidBrush & \XSolidBrush  & 3.84 & -\\
            Dance w/Melody~\cite{tang2018dance} & \Checkmark/\XSolidBrush & 21 & \XSolidBrush & \XSolidBrush & 4 & \Checkmark & \Checkmark & \XSolidBrush & \XSolidBrush & 1.6 & 92.5\\
            Music2Dance~\cite{zhuang2022music2dance} & \Checkmark/\XSolidBrush & 55 & \Checkmark & \XSolidBrush & 2 & \Checkmark & \Checkmark & \XSolidBrush & \XSolidBrush & 0.96 & -\\
            AIOZ-GDANCE~\cite{le2023music}& \Checkmark/\Checkmark & 24 & \XSolidBrush & \XSolidBrush & 7 & \XSolidBrush & \XSolidBrush & \Checkmark & \Checkmark & \textbf{16.7} & 73.8\\
            PhantomDance~\cite{li2022danceformer} & \Checkmark/\XSolidBrush & 24 & \XSolidBrush & \XSolidBrush & 13 & \XSolidBrush & \Checkmark & \XSolidBrush & \XSolidBrush &  9.6 & 133.3\\
            AIST++~\cite{li2021ai} & \Checkmark/\Checkmark & 24 & \XSolidBrush & \XSolidBrush & 10 & \XSolidBrush & \Checkmark & \XSolidBrush & \Checkmark & 5.2 & 13.3\\
            MMD~\cite{chen2021choreomaster} & \Checkmark/\Checkmark & 52 & \Checkmark & \XSolidBrush & 4  & \Checkmark & \Checkmark & \XSolidBrush & \XSolidBrush &  9.9 & -\\
            FineDance~\cite{li2023finedance}  &  \Checkmark/\Checkmark & 52 & \Checkmark & \XSolidBrush & \textbf{22} & \Checkmark & \Checkmark & \XSolidBrush & \Checkmark & 14.6 & 152.3\\
    
            \midrule [0.5pt]
            \textbf{SoulDance (Ours)} &  \Checkmark/\Checkmark & \textbf{55+\underline{\textit{52}}} & \Checkmark & \Checkmark & 15  & \Checkmark & \Checkmark & \Checkmark & \Checkmark & 12.5 & \textbf{158.5}\\
            \bottomrule [1pt] 
        \end{tabular}
    }
    \vspace{0.1mm}
    \caption{\textbf{Comparisons of 3D Dance Datasets}. ``Pos" and ``Rot" represent 3D position and rotation information, respectively. ``\underline{\textit{52}}" refers to the parameters associated with ARKit blendshapes. ``avg Sec per Seq" indicates the average duration in seconds for each sequence.}
    \label{tab:data_compare}   
    \vspace{-1em}
\end{table*}

\section{SoulDance Dataset}
\label{sec:souldance}
\textit{SoulDance} is the first high-quality captured 3D dance dataset that incorporates expressive facial movements in music-dance pairs. As shown in Table~\ref{tab:data_compare}, most existing datasets lack detailed hand movements and almost entirely omit facial expressions, making it challenging to generate truly holistic dance from such data. Most dance datasets rely on pose estimation methods~\cite{sun2022dancetrack, fang2017rmpe} to extract dance movements from videos~\cite{li2021ai, li2022danceformer, le2023music}. However, these pipelines often yield low-quality data, as accurate gesture movements and expressions are hard to obtain from video sources alone. To overcome these limitations, we employed a professional holistic motion capture system to record dance movements, ensuring precise synchronization of the body, hands, and face in both temporal and spatial domains, thereby faithfully preserving real-world motion dynamics. 



\noindent\textbf{Holistic Motion Capture System.}
To capture high-quality and accurate dance motions, we constructed a comprehensive motion capture system, as illustrated in Figure~\ref{fig:teaser_image}. The system consists of two main components. First, we use a marker-based professional motion capture system~\cite{chingmu} to capture body and hand movements, storing the data in BVH format. Second, we developed a stable facial expression capture setup using an iPhone 12 and a helmet (dashed oval in Figure~\ref{fig:teaser_image}), utilizing ARKit~\cite{baruch2021arkitscenes} to record face blendshapes with 52 parameters. In a 140 $m^2$ motion capture studio, we use 15 cameras to record the dance movements. A professional choreographer arranges the dance according to the music, and performers rehearse in a regular dance studio. Once proficient, they move to the professional motion capture studio, where they wear motion capture suits for the recording. The motion capture software CMAvatar~\cite{chingmu} is used, with a recording frame rate of 60 FPS. 
After data collection, the raw motion data for the body and hands is refined and retargeted to the SMPL-X~\cite{SMPL-X:2019} model, as described in Appendix~\ref{supl:refine}. Meanwhile, the raw ARKit Face Blendshape data is converted into FLAME~\cite{FLAME:SiggraphAsia2017} parameters compatible with SMPL-X (see Appendix~\ref{supl:flame} for details).

\noindent\textbf{High-Quality Dance-Music Pairs.} We enlisted five renowned professional dancers to perform, ensuring that the dance movements are executed with skill and expertise. Throughout the entire recording process, a director supervised the quality of dance and their alignment with the music, while a engineer was on hand to adjust and correct settings in case of recording misalignment. Furthermore, expert choreography was incorporated to guarantee perfect alignment between music and dance. This meticulous process resulted in a high-quality, temporally and spatially coherent dance motion-music dataset, providing a solid foundation for expressive and coordinated dance generation. Additionally, we provide an analysis and visual demonstration of the \textit{SoulDance} dataset. For more details, please refer to Appendix~\ref{supl:add_dataset}.


\begin{figure}[h]
	\centering
	\includegraphics[width=\linewidth]{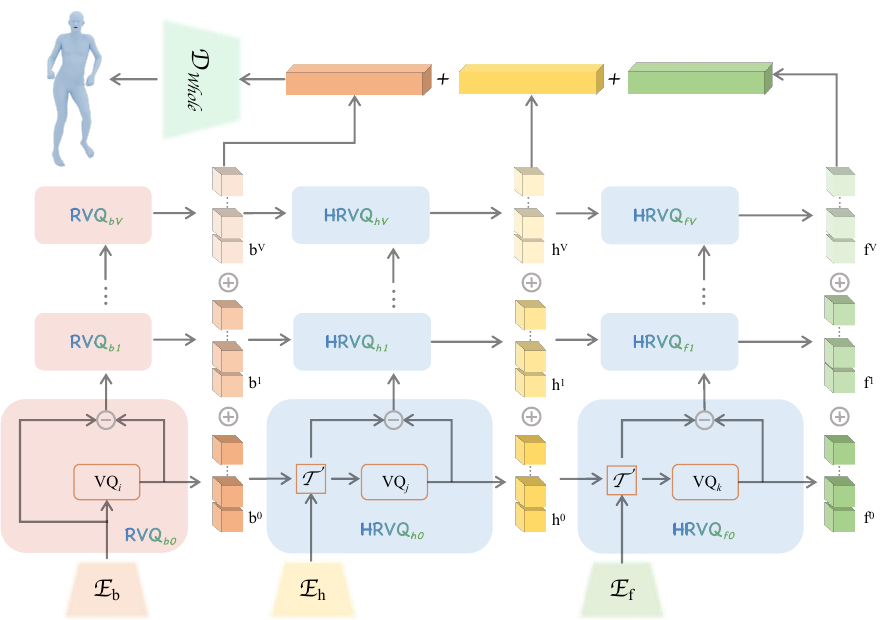}
        \vspace{-1em}
	\caption{\textbf{Hierarchical Residual Vector Quantization.} For the body component, we apply $\mathit{RVQ}_{bv}$. In this process, each quantized layer output $b^v$ serves as a body hint that is passed sequentially to the hands component, which is then quantized using $\mathit{HRVQ}_{hv}$. Similarly, the face component undergoes $\mathit{HRVQ}_{fv}$, using hands hints derived from the preceding quantization steps to guide the encoding process.}
        \label{fig:hrvq}
        \vspace{-1.5em}
\end{figure}

\section{SoulNet}
\label{sec:soulnet}


Given a music $C$, our goal is to generate high-quality and music-aligned holistic 3D dance motions $\tilde m_{1:N}$, where $\tilde m_i \in \mathbb{R}^D$, with $D$ representing the pose feature dimension and length $N$ determined by the duration of $C$.

\begin{figure*}[h]
	\centering
	\includegraphics[width=\linewidth]{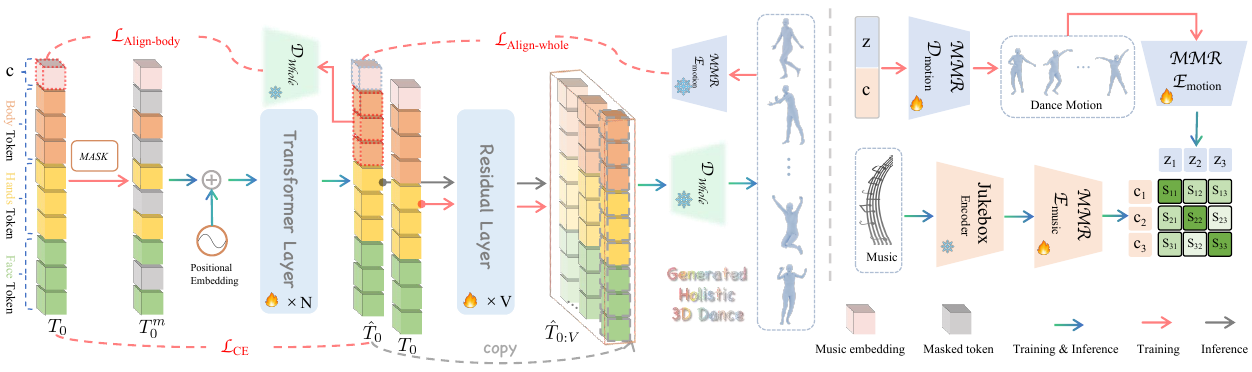}
	\caption{\textbf{An overview of our \textit{SoulNet} framework.} On the left, the Music-Aligned Generative Model (MAGM) Training and Inference consists of two stages: (1) transformer layers generate primary motion tokens in the base layer, and (2) residual layers refine motion through $V-1$ quantization layers. On the right, the Music-Motion Retrieval Module (MMR) illustrates the motion-music alignment process during its pretraining. Once trained, MMR supervises the training of MAGM through $\mathcal{L}_{\text{Align-body}}$   and $\mathcal{L}_{\text{Align-whole}}$ , ensuring effective alignment between generated motion and music.}
	\label{fig:gpt}
	\vspace{-1.5em}
\end{figure*}

We propose \textit{SoulNet} and the overview of the holistic dance generation framework is shown in Figure~\ref{fig:gpt}. We first employ Hierarchical Residual Vector Quantization (HRVQ) to encode and tokenize the spatially structured dance sequence into a multi-layer codebook. This tokenization efficiently constrains the dance motion space and captures the prior relationships among the body, hands, and face (Section~\ref{subsec:hrvq}). Next, we use the Music-Aligned Generative Model (MAGM), which leverages transformer Layers to model the dance token sequence and employs residual layers to refine tdance details (Section~\ref{subsec:magm}). Meanwhile, a pre-trained Music-Motion Retrieval (MMR) model serves as a prior to align the generated dance sequence with the music (Section~\ref{subsec:mmr}). Furthermore, MMR provides additional constraints on the dance motion space, further reducing the training difficulty of MAGM and improving the quality of the generated dance.


\subsection{Hierarchical Residual Vector Quantization}
\label{subsec:hrvq}
Previous works~\cite{siyao2022bailando, guo2022tm2t, jiang2023motiongpt} leverage VQ-VAEs to quantize dance sequences $m_{1:N} \in \mathbb{R}^{N \times D}$ into a sequence of discrete motion tokens $z_{1:N} \in \mathbb{R}^{n \times d}$ with downsampling ratio of $n/N $ and latent dimension $d$ by constructing a reusable codebook $\mathcal{C} = \{c_k\}_{k=1}^{K} \subset \mathbb{R}^{d}$ in an unsupervised manner. This process can be described as following:
\begin{equation}
\begin{aligned}
\hat{m}_{1:N} = \mathcal{D}(VQ(\mathcal{E}(m_{1:N}))),
\end{aligned}
\end{equation}
where $\mathcal{E}$ encodes $m_{1:N}$ into latent vector, $VQ(\cdot)$ represents quantization, and $\mathcal{D}$ projects the quantized code sequence back into the motion to obtain the reconstructed dance $\hat{m}_{1:N}$. However, $VQ(\cdot)$ inevitably results in information loss, which limits reconstruction quality.
Recent works~\cite{zeghidour2021soundstream, yao2024moconvq, guo2024momask} introduce RVQ, which iteratively quantizes the residual error at each level from the previous one. 

To leverage the multi-layer encoding capability of RVQ for capturing complex and holistic dance movements, we introduce HRVQ, which is designed with a multi-layered body-hand-face chain that learns dependencies across the body, hands, and facial expressions, as illustrated in Fig.~\ref{fig:hrvq}. Specifically, we begin by decomposing the holistic dance sequence $m_{1:N}$ into three components: body motion $m_{1:N}^b$, hands motion $m_{1:N}^h$, and face motion $m_{1:N}^f$. These components are then processed through their respective encoders $\mathcal{E}$ to produce primary quantization vectors $b^0$, $h^0$, and $f^0$. 
Next, we iteratively quantize the residual $r_b^v$ of the body motion component, yielding the quantization vector $b^v$ for each layer $v = 0, \dots, V$, where $V$ denotes the total number of layers. Specially, $b^0$ serves as the primary vector, and subsequent vectors $b^{1:V}$ capture residuals. This iterative structure 
enables progressively finer quantization and efficient encoding across multiple levels. Subsequently, we apply a transformation process, $\mathcal{T}$, to combine the hand motion residual $r_h^v$ with $b^v$, then quantize the result to obtain $h^v$. The same process applies for the face motion, yielding $f^v$. Formally, this can be expressed as:
\begin{equation}
    \label{eq:hrvq_z}
    \left \{
    \begin{aligned}
        & b^v=VQ_b(r^v_b) \\
        & h^v=VQ_h(\mathcal{T}(r^v_h, b^v)) \\
        & f^v=VQ_f(\mathcal{T}(r^v_f, h^v)) ,\\
    \end{aligned}
    \right.
\end{equation}
where $VQ_b$, $VQ_h$, and $VQ_f$ represent the quantization functions for body, hands, and face, respectively. The next residuals are then represented as:
\begin{equation}
    \label{eq:hrvq}
    \left \{
    \begin{aligned}
        & r^0_b=\mathcal{E}_b(m_{1:N}^b), r^{v+1}_b = r^v_b - b^v \\
        & r^0_h=\mathcal{T}(\mathcal{E}_h(m_{1:N}^h), b^0), r^{v+1}_h = r^v_h - h^v \\
        & r^0_f=\mathcal{T}(\mathcal{E}_f(m_{1:N}^f), h^0), r^{v+1}_f = r^v_f - f^v, \\
    \end{aligned}
    \right.
\end{equation}
where $\mathcal{E}_b$, $\mathcal{E}_h$, and $\mathcal{E}_f$ denote the encoders for the body, hands, and face, respectively.
After applying HRVQ, the motion reconstruction is given by:
\begin{equation}
\hat{m}_{1:N} = \mathcal{D}_{\text{whole}}\left(\text{concat}\left(\sum_{v=0}^{V} b^v, \sum_{v=0}^{V} h^v, \sum_{v=0}^{V} f^v\right)\right),
\end{equation}
where $ \mathcal{D}_{\text{whole}} $ is the motion decoder that projects the latent sequence back into the motion space. This formulation effectively combines reconstructed components from the body, hands, and face to yield the final motion sequence $ \hat{m}_{1:N} $.
Finally, the hierarchical residual VQ-VAE is trained via the motion reconstruction loss combined with the latent embedding loss at each quantization layer for every part:
\begin{equation}
\begin{aligned}
\mathcal{L}_{hrvq} = & \left \| m - \hat{m} \right \|_1 + \alpha \sum_{v=1}^{V}\left \| r^v_b - sg[b^v] \right \|_2 \\
& + \beta \sum_{v=1}^{V}\left \| r^v_h - sg[h^v] \right \|_2 + \gamma \sum_{v=1}^{V}\left \| r^v_f - sg[f^v] \right \|_2 ,
\end{aligned}
\end{equation}
where $sg[\cdot]$ denotes the stop-gradient operation. The parameters $\alpha$, $\beta$, and $\gamma$ are factors for the embedding constraints. Since HRVQ relies on discrete VQ token sampling, we adopt the \textit{Gumbel-Softmax}~\cite{jang2016categorical} trick to enable gradient backpropagation.


\subsection{Music-Aligned Generative Model}
\label{subsec:magm}

After HRVQ quantization, the body, hands, and face tokens are concatenated to form unified dance holistic tokens, $T_v = \text{concat}(b^v, h^v, f^v)$ for $v=0, \dots, V$. 
To generate diverse and coordinated holistic dance movements from the token sequence obtained via HRVQ, two key requirements must be satisfied: (1) The model needs to effectively capture the relationships between tokens, preserving their details and hierarchical structure. (2) The token sequence must be aligned with the input music to ensure the generated dance corresponds to the rhythm and style of the music.

To address these problems, We propose Music-Aligned Generative Model (MAGM), a network designed to generate holistic dance motions conditioned on a music $C$, as illustrated in Figure~\ref{fig:gpt}. We employ the MMR music encoder (in Section ~\ref{subsec:mmr}) to extract music features, appending the music feature to $T_0$, and applying random mask and positional encoding before passing it through a Transformer layer $\theta$ to predict masked tokens, resulting in $\hat{T}_0$. Mathematically, $\theta$ is optimized as follows:
\begin{equation}
\mathcal{L}_{\text{mask}} = \sum_{t_i \in \text{mask}} -\log_{\theta} (T_0 | T_0^m, C) + \lambda_b \mathcal{L}_{\text{Align-body}},
\end{equation}
where $T_0 = \{ t_i \}_{i=1}^n$. Given $C$ and $T_0$ as conditions, we use $V-1$ residual layers $\phi$ to predict residuals $T_{1:V}$ using a loss function:
\begin{equation}
\mathcal{L}_{\text{res}} = \sum_{i=1}^{V} -\log_{\phi} (T_{i:V} | T_0, C) + 
\lambda_w \mathcal{L}_{\text{Align-whole}},
\end{equation}
$\lambda_b$ and $\lambda_w$ are hyperparameters, both set to 0.5. 
$\hat{T}_0$ alone is insufficient to generate fine-grained dance, so we use $V$ residual layers to predict the residuals $\hat{T}_{1:V}$. These are concatenated with $\hat{T}_0$ to form $\hat{T}_{0:V}$, which is decoded by $\mathcal{D}_{\text{whole}}$ to generate the final motion.
The dance-music alignment priors for both $\mathcal{L}_{\text{Align-whole}}$ and $\mathcal{L}_{\text{Align-body}}$ are derived from the Music-Motion Retrieval Module.

\subsection{Music-Motion Retrieval Module}
\label{subsec:mmr}
Establishing precise alignment between music and dance motion represents a fundamental challenge in music-conditioned dance generation. While recent advances in cross-modal generation domains such as text-to-image synthesis~\cite{rombach2022high, saharia2022photorealistic} have demonstrated the importance of well-aligned embedding spaces through models like CLIP~\cite{radford2021learning}, music-dance alignment remains comparatively underexplored. Current dance generation approaches predominantly extract low-level acoustic features using conventional tools like Librosa~\cite{jin2017towards} or pretrained music encoders such as Jukebox~\cite{dhariwal2020jukebox}, and directly feed these features to motion generators without explicit cross-modal alignment. This direct projection approach creates a significant modality gap between audio signals and motion representations, resulting in dance sequences that often lack precise synchronization with musical elements.

To address this limitation, we propose the Music-Motion Retrieval (MMR) module, a cross-modal alignment framework designed specifically for the music-dance domain. As illustrated in Figure~\ref{fig:gpt} (right), our architecture consists of three principal components: (1) \textbf{Motion Encoding}, where input motion sequences $M \in \mathbb{R}^{T \times D_m}$ are compressed into latent codes $\mathbf{z} = \{z_i\}_{i=1}^d$ through a temporal encoder $\mathcal{E}_{\text{motion}}$ with learnable downsampling operations. (2) \textbf{Music Processing}, in which raw audio inputs $C$ are first encoded using the pretrained Jukebox encoder~\cite{dhariwal2020jukebox}, followed by a dedicated music encoder $\mathcal{E}_{\text{music}}$ that projects features into a motion-aligned latent space $\mathbf{c} = \{m_i\}_{i=1}^d$. (3) \textbf{Cross-Modal Fusion}, where the aligned latent representations $\mathbf{z}$ and $\mathbf{c}$ are concatenated and decoded through a transformer-based motion decoder $\mathcal{D}_{\text{motion}}$ to reconstruct coherent dance sequences. Our encoder-decoder architecture extends the TEMOS framework~\cite{petrovich2022temos} with modality-specific adaptations. 
We apply $\mathcal{L}_{\text{Align-body}}$ to guide the generation of bodily structure, and employ $\mathcal{L}_{\text{Align-whole}}$ to align the whole-body motion with musical affect. Both losses are derived from an alignment objective:
\begin{equation}
\small
\begin{split} 
    \mathcal{L}_{\text{Align}} &= - \frac{1}{2N} \sum_{i} \left( \log \frac{e^{S_{ii}}}{\sum_{j} e^{S_{ij}}} + \log \frac{e^{S_{ii}}}{\sum_{j} e^{S_{ji}}} \right) ~,
\end{split}
\label{eq:infonce}
\end{equation}
The difference arises from the similarity matrix $S$, as it is derived from different motion representations.


During inference, the music signal $ C $ and motion sequences $ M \in \mathbb{R}^{T \times D_m} $ are encoded into latent representations $ \mathbf{z} $ and $ \mathbf{c} $ via the MMR encoder. 
In the MAGM dance generation framework, $ \mathcal{L}_{\text{Align-body}} $ enforces local temporal alignment between body movements and musical beats, while $ \mathcal{L}_{\text{Align-whole}} $ regulates global synchronization of holistic motion dynamics with musical affect. To achieve this dual granularity, we train two specialized MMR modules using two different datasets. Please refer to Appendix~\ref{supl:traing_mmr}.

\section{Experiments}
\label{sec:experiment}

\noindent\textbf{Dance datasets.} The AIST++ dataset~\cite{li2021ai} is widely used in 3D dance, containing 1,408 sequences and a total of 5.2 hours of dance with music. The FineDance dataset~\cite{li2023finedance}, captured using an optical motion capture system, provides 7.7 hours of dance data, consisting of 831,600 frames. For fairness in our experiments, we randomly selected 6.3 hours of data from the \textit{SoulDance} dataset for experiment. Additionally, all dance datasets are represented using our specified approach (see Appendix~\ref{supl:motion_rep} for details). Finally, all datasets are split into training, testing, and validation sets with an 8:1:1 ratio. The implementation details of our method, \textit{SoulNet}, can be found in Appendix~\ref{supl:soulnet_impl}.

\begin{table}
    \resizebox{\linewidth}{!}{
        \begin{tabular}{lcccccccc}
            \toprule [1pt]
            \noalign{\smallskip}
            \multirow{2}{*}{Method}& \multicolumn{4}{c}{SoulDance Dataset} & \multicolumn{3}{c}{FineDance Dataset} & \multicolumn{1}{c}{AIST++ Dataset} \\  \cmidrule(lr){2-5} \cmidrule(lr){6-8} \cmidrule(lr){9-9}
            \noalign{\smallskip}
            & $all\downarrow$ & $body\downarrow$ & $hand\downarrow$ & $face\downarrow$ & $all\downarrow$ & $body\downarrow$ & $hand\downarrow$ & $body\downarrow$ \\
            \noalign{\smallskip}\midrule [0.5pt]
            \noalign{\smallskip}
            Vanilla VQ (512) & 137.130 & 97.660 & 129.518  &4.013 &	137.646	& 96.598 & 136.329 & 42.513\\
            Vanilla VQ (1024) & 136.752 & 95.809 & 128.628 & 3.739 & 136.785 & 97.328 & 135.194 & 41.369\\
            RVQ-5 & 108.983 & 71.831 & 106.836 &  2.379 & 98.399 & 59.281 & 102.706 & 24.246 \\
            HRVQ-5 (Ours) & \textbf{83.679} & \textbf{47.895} & \textbf{85.085} & \textbf{1.153} & \textbf{90.705} &\textbf{51.708} & \textbf{95.563} & \textbf{24.246} \\
            \bottomrule [1pt] 
            \noalign{\smallskip}
    \end{tabular}}
    \caption{\textbf{Comparison of the motion reconstruction. }We evaluate the motion reconstruction performance of various quantization methods using MPJPE (measured in mm). We use Face Vertex Error (measured in $10^{-4}$ mm) to evaluate the reconstruction quality of facial motion. See Appendix~\ref{supl:rec_eval} for more details}
    \label{tab:vqs_compare} 
    \vspace{-1em}
\end{table}

\noindent\textbf{Baselines.} We compare our method with several state-of-the-art dance generation methods. FACT~\cite{li2021ai} is an autoregressive dance generation method that is introduced alongside the AIST++ dataset. Bailando~\cite{siyao2022bailando} is a transformer-based generation method that firstly incorporate VQ-VAE for dance motion generation. EDGE~\cite{tseng2023edge} is a diffusion-based dance generation algorithm, achieving high-quality body-only dance generation. FineNet~\cite{li2023finedance} is the first approach to generate motions that include both body and hand movements, introduced alongside the FineDance dataset.


\begin{table}[t]
    \resizebox{\linewidth}{!}{
        \begin{tabular}{lcccccccc}
            \toprule [1pt]
            \noalign{\smallskip}
            \multirow{2}{*}{Method}& \multicolumn{4}{c}{HRVQ} & \multicolumn{4}{c}{RVQ} \\  \cmidrule(lr){2-5} \cmidrule(lr){6-9}
            \noalign{\smallskip}
            & $all\downarrow$ & $body\downarrow$ & $hand\downarrow$ & $face\downarrow$ & $all\downarrow$ & $body\downarrow$ & $hand\downarrow$ & $face\downarrow$ \\
            \noalign{\smallskip}\midrule [0.5pt]
            \noalign{\smallskip}
            V=1 & 104.859 & 67.470 & 103.092 &1.990 &126.071&86.234&120.795&3.224\\
            V=2 & 98.180 & 61.156 & 97.365 & 1.694 &121.383&81.795&116.148&2.856 \\
            V=3 & 92.777 & 55.533 & 93.776 & 1.502 &111.660&78.883&106.972&2.618 \\
            V=4 & 85.035 & 51.360 & 85.973 & 1.291 &113.503&72.570&111.069&2.475\\
            V=5 & 83.679 & 47.895 & 85.085 & 1.153 &108.983&71.831&106.836&2.379\\
            V=6 & \textbf{79.518} & 46.806 & \textbf{80.015} & 1.117 & 105.846 &65.947&105.208&2.223 \\
            V=7 & 84.080 & \textbf{46.573} & 86.218 & \textbf{1.007} & \textbf{101.209}&\textbf{63.457}&\textbf{99.794}&\textbf{2.098} \\
            \bottomrule [1pt] 
            \noalign{\smallskip}
    \end{tabular}}
    \caption{\textbf{Effect of Hierarchical Residual Layers.} We vary the number of residual layers from 1 to 7 to investigate the impact of different $V$ values on motion reconstruction quality.}
    \label{tab:diff_hrvq_v} 
    \vspace{-1.5em}
\end{table}

\begin{table*}
    \resizebox{\linewidth}{!}{
        \begin{tabular}{lcccccccccccccc}
            \toprule [1pt]
            \multirow{2}{*}{Method}&
            \multicolumn{8}{c}{SoulDance Dataset} & \multicolumn{6}{c}{AIST++ Dataset} \\  \cmidrule(lr){2-9} \cmidrule(lr){10-15}
            \noalign{\smallskip}
            & FID $\downarrow $& FID$_h$ $\downarrow$ & Div $\uparrow$ & Div$_h$ $\uparrow $& MM $\uparrow $& MMR-MS $\downarrow$ & BAS $\uparrow $& EAS $\uparrow$ & FID $\downarrow $& Div $\uparrow$ & MM $\uparrow $&MMR-MS $\downarrow$ & BAS $\uparrow $&Run Time $\downarrow $ \\
            \noalign{\smallskip}\midrule[0.5pt]
            \noalign{\smallskip}
            Ground Truth & - & - & 1.322  &5.713 &-& 0.319 & 0.253 & -& - & 0.663 & - & 0.486 & 0.237 & -\\
            \midrule [0.5pt]
            FACT~\cite{li2021ai} & 1.008 & 0.408 & 0.646  &0.799 &0.656& 0.685 & 0.221 & 0.358 & 0.138 & 0.654 & 0.722 & 0.702 & 0.213 & 1.782\\
            Bailando~\cite{siyao2022bailando} & 1.379 & 2.244 & 1.307  &2.126 &1.117& 0.585 & 0.236 & 0.401 & 11.079 & 3.512 & \textbf{3.071} & 0.707 & 0.229 & 0.289\\
            EDGE~\cite{tseng2023edge} & 2.619 & 3.694 & 0.723  &4.025 &0.745& 0.716 & 0.241 & 0.246 & 21.370 & 1.562 & 1.397 & 0.687 & 0.233 & 1.521\\
            FineNet~\cite{li2023finedance} & 1.463 & 2.524 & 1.262  &3.448 &0.832& 0.694 & 0.213 & 0.263 & 3.127 & \textbf{3.836} & 0.885 & 0.696 & 0.223 & 1.637\\
            \midrule [0.5pt]
            \textbf{SoulNet (w/o. MMR)} & 0.048 & 1.454 & 1.303  &\textbf{4.178} &1.289& 0.418 & 0.240 & 0.435 & 0.206 & 0.698 & 0.855 & 0.714 & 0.232 & 0.086\\
            \textbf{SoulNet} & \textbf{0.029} & \textbf{0.375} & \textbf{1.312}  &3.423 &\textbf{1.310}& \textbf{0.369} & \textbf{0.244} & \textbf{0.594} & \textbf{0.081} & 1.649 & 1.359 & \textbf{0.580} & \textbf{0.242} & \textbf{0.086} \\
            \bottomrule [1pt] 
        \end{tabular}
        }
    \caption{\textbf{Comparison with Different Methods.} We compare various dance generation methods on the \textit{SoulDance} and AIST++ datasets, and the results show that \textit{SoulNet} achieves state-of-the-art performance on \textit{SoulDance} dataset. However, due to the small size of AIST++ dataset, SoulNet shows slightly lower diversity compared to other methods.}
    \label{tab:sotas_compare} 
    \vspace{-1em}
\end{table*}

\begin{table}
    \resizebox{\linewidth}{!}{
        \begin{tabular}{lccccc}
            \toprule [1pt]
            Method
            & FID $\downarrow$ & Div $\uparrow$ & MM $\uparrow$ & BAS $\uparrow$ & MMR-MS $\downarrow$ \\
            \noalign{\smallskip}\midrule[0.5pt]
            \noalign{\smallskip}
            VQ-512 & 1.610 & 0.540 & 0.509  &0.237 &0.703\\
            VQ-512 + MMR & 1.552 & 0.563 & 0.576  &0.241 &0.697\\
            RVQ-512 & 0.082 & 1.308 & 1.233  &0.216 &0.571\\
            RVQ-512 + MMR & 0.067 & \textbf{1.329} & 1.250  &0.242 &0.540\\
            RVQ-1024 + MMR & 0.090 & 1.182 & 1.228  &0.234 &0.603\\
            HRVQ-512 & 0.048 & 1.303 & 1.289  &0.240 &0.418\\
            HRVQ-512 + MMR & \textbf{0.029} & 1.312 & \textbf{1.310}  &\textbf{0.244} &\textbf{0.369}\\
            \bottomrule [1pt] 
        \end{tabular}
        }
        \caption{\textbf{Ablation Study of HRVQ and MMR.} We conduct experiments on HRVQ with codebook sizes of 512 and 1024, comparing the results with and without MMR supervision.}
    \label{tab:abl_soulnet} 
    \vspace{-1em}
\end{table}

\begin{table}[ht]
\centering
\resizebox{\linewidth}{!}{
	\begin{tabular}{lccccccc}
	\toprule [1pt] 
	\multicolumn{2}{c}{Ablations} & \multicolumn{5}{c}{Metrics} \\ \cmidrule(lr){1-2}   \cmidrule(lr){3-7}   
        $\mathcal{L}_{\text{Align-body}}$  & $\mathcal{L}_{\text{Align-whole}}$ & FID $\downarrow$ & Div $\uparrow$ & MM $\uparrow$ & BAS $\uparrow$ & MMR-MS $\downarrow$ \\
	\noalign{\smallskip}\hline\noalign{\smallskip}
         \checkmark  & \checkmark &  \textbf{0.029} & 1.312 & 1.310  &\textbf{0.244} &\textbf{0.369}\\
         \checkmark  &   &0.031  &1.331  &1.327&0.242&0.387\\
         &  \checkmark &0.042  &\textbf{1.344} &\textbf{1.338}&0.237&0.372\\
		\bottomrule [1pt] 
               \noalign{\smallskip}
	\end{tabular}}
	\caption{\textbf{Ablation Study of MMR.} We conduct experiments on the \textit{SoulDance} dataset to compare the effects of $\mathcal{L}_{\text{Align-body}}$ and $\mathcal{L}_{\text{Align-whole}}$ on dance motion alignment.}
	\label{tab:abl_alignloss}   
    \vspace{-1.5em}
\end{table}

\subsection{Comparison to SOTA Methods}
In this section, we compare our \textit{SoulNet} to SOTA methods on (1) our proposed MMR-Matching Score, (2) our proposed Emotion Alignment Score, (3) public benchmarks, and (4) user study

\noindent\textbf{MMR-Matching Score.} To quantify the alignment between music and the generated dance, we employ MMR encoders to project both modalities into latent space. The MMR-Matching Score (MMR-MS) is then computed as the Euclidean distance between these latent representations:
\begin{equation}
\text{\textit{MMR-MS}} = \sqrt{\mu \cdot\sum_{i=1}^{d} (\mathbf{z}_i - \mathbf{m}_i)^2+\lambda \cdot \sum_{t=2}^T \left\Vert \Delta\mathbf{z}^t - \Delta\mathbf{m}^t \right\Vert_2}
\end{equation}
where the first term measures the feature space distance, the second term quantifies the distance in feature dynamics. Specifically, $ \mathbf{z} = \{z_i\}_{i=1}^d $ and $ \mathbf{c} = \{m_i\}_{i=1}^d $ represent the latent representations of motion and music, respectively. To compute the second term, we first segment both dance and music sequences into non-overlapping 1-second segments, resulting in a set of $T$ sequences. These segments are then encoded using MMR encoders to obtain their latents. The temporal differences between consecutive latents are defined as $\Delta\mathbf{z}^t = \mathbf{z}^t - \mathbf{z}^{t-1}$ and $\quad \Delta\mathbf{m}^t = \mathbf{m}^t - \mathbf{m}^{t-1}$ that quantify the temporal variations in the latent space for the dance and music, respectively. Typically, we set $\mu=0.7, \lambda=0.3$. As shown in Table~\ref{tab:sotas_compare}, our results demonstrate that \textit{SoulNet} achieves superior music-motion alignment, significantly outperforming existing approaches.

\noindent\textbf{Emotion Alignment Score.} Expressions that align well with the emotional tone of the music create a more immersive and compelling experience. However, existing methods lack an explicit evaluation of how well generated facial expressions align with music. To bridge this gap, we introduce the Emotion Alignment Score (EAS) that quantifies the alignment between generated facial expressions and the emotional tone of the music. To represent facial emotions, we adopt a 7-expression model based on Ekman’s theory~\cite{ekman1971constants}. For expression prediction, we leverage state-of-the-art facial expression recognition algorithms~\cite{savchenko2023facial, xue2022vision} to classify the generated expressions and compute alignment accuracy. We directly use the accuracy of comparing predicted expressions with ground truth expressions as the Emotion Alignment Score.
As shown in Table~\ref{tab:sotas_compare}, higher EAS indicate that our method achieves superior alignment between facial expressions and the emotional tone of the music.

\begin{figure}[h]
	\centering
	\includegraphics[width=\linewidth]{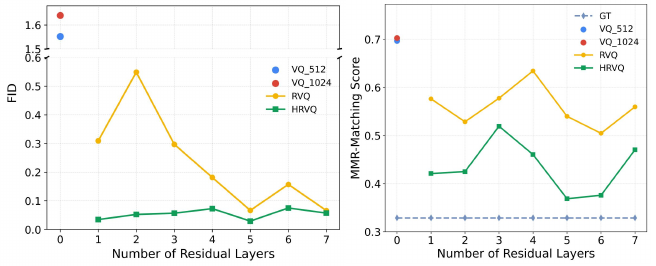}
        \vspace{-1em}
	\caption{\textbf{FID and MMR-Matching Score.} Performance comparison across varying numbers of residual layers for different quantization methods.}
        \label{fig:abl_fid_mmr}
        \vspace{-1.5em}
\end{figure}

\noindent\textbf{Public benchmarks.} (1) \textit{FID} score. Fréchet inception distance (FID) is widely used to measure how close the distribution of the generated dances is to that of the ground truth. We follow the approach introduced in HumanML3D~\cite{guo2022generating}, which is widely adopted in motion generation studies for assessing motion quality~\cite{guo2024momask, lu2023humantomato, guo2022tm2t, li2023finedance}. (2) \textit{Diversity}. Diversity evaluate the average feature distance between generated dances for different input music. The same feature extractor used in FID is used again. (3) \textit{Hand FID} score and \textit{Hand Diversity}. Similarly, we extract hand motion features and calculate the FID and diversity for hand motion. (4) \textit{Multimodality}. We follow Guo \textit{et al.}\cite{guo2022generating} to evaluate the average feature distance between the 10 choreography versions of every music. This metric measures the model's ability to generate different dances for the same music. (5) \textit{Run Time}. We evaluated the average runtime to generate dance sequences of equal length during the inference process. Qualitative results further demonstrate that our method, \textit{SoulNet}, achieves high-quality and diverse holistic dance generation. Please refer to Appendix~\ref{supl:add_vis_res}.

\noindent\textbf{User Study.} In the user study, 28 participants—including 7 professional dancers—evaluated 22 pairs of videos. The durations of videos A and B ranged from 5 to 8 seconds. Participants were asked to answer a series of questions (see Appendix~\ref{supl:user_study_details} for details), each rated on a scale from 1 to 10. The scores were denoted as Whole Score (WS), Body Score (BS), Hands Score (HS), Emotion Score (ES), and Alignment Score (AS), corresponding to the respective questions in order. The results (Table~\ref{tab:user_study}) demonstrate that our proposed \textit{SoulDance} dataset and \textit{SoulNet} model outperform existing datasets and methods across all evaluation metrics.

\subsection{Ablation Study}
\noindent\textbf{Different VQ-based Approaches.} In Table~\ref{tab:vqs_compare}, we provide a comprehensive evaluation of various VQ-based methods and their impact on motion reconstruction. Our proposed HRVQ demonstrates SOTA performance, achieving substantially lower reconstruction errors comparing to VQ and RVQ quantization techniques, and qualitative results of the reconstruction can be found in Appendix~\ref{supl:add_vis_res}. In Table~\ref{tab:abl_soulnet}, we present the performance of different VQ-based approaches on dance motion generation in the SoulDance dataset. Qualitative results also clearly favor our method over others in Figure~\ref{fig:soulnet_abl}. Overall, HRVQ consistently outperforms in both motion reconstruction and dance generation tasks, establishing itself as a leading method in this domain.

\noindent\textbf{Effect of Quantization Layers.}
In Table~\ref{tab:diff_hrvq_v}, we analyze the impact of different quantization layer counts $V$ for both HRVQ and RVQ methods. Generally, adding more residual VQ layers enhances reconstruction precision, though this also increases the computational load on the MAGM for dance token generation. As shown in Figure~\ref{fig:abl_fid_mmr}, we observe a noticeable decline in generation quality when the number of residual layers exceeds 5 or 6, both in RVQ and HRVQ. Therefore, to balance reconstruction accuracy and generation efficiency, \textit{SoulNet} is designed with $V=5$ (i.e., $5$ residual VQ layers) for HRVQ. 

\noindent\textbf{Music-Aligned Strategy.} To evaluate the effectiveness of the proposed Music-Aligned strategy, we compare the generation results with and without the MMR module. As shown in Table~\ref{tab:abl_soulnet}, incorporating the MMR module enhances the quality of the generated dance and improves alignment with the music. Moreover, this enhances the qualitative effects and additional details can be found in Appendix~\ref{supl:add_vis_res}. 
To further analyze the contributions of each alignment loss, we performed ablation studies on $\mathcal{L}_{\text{Align-body}}$ and $\mathcal{L}_{\text{Align-whole}}$. As shown in Table~\ref{tab:abl_alignloss}, $\mathcal{L}_{\text{Align-body}}$ significantly improves local feature alignment (FID and BAS), while $\mathcal{L}_{\text{Align-whole}}$ enhances global structure alignment (MMR-MS). 

\begin{table}
\centering
    \resizebox{0.8\linewidth}{!}{
        \begin{tabular}{lccccc}
            \toprule [1pt]
            Dataset / Method
            & WS  & BS  & HS & ES & AS \\
            \noalign{\smallskip}\midrule[0.5pt]
            \noalign{\smallskip}
            AIST++ Dataset & 4.44 &5.22 &4.67&3.56&3.89\\
            FineDance Dataset & 6.89 & 6.57 &6.44&5.44 &6.67\\
            \textbf{SoulDance Dataset} & \textbf{8.56} & \textbf{8.67}&\textbf{8.22}&7.56 &\textbf{8.44} \\
            \midrule [0.5pt]
            FACT~\cite{li2021ai} & 5.00 & 5.11&5.22&5.33 &4.44 \\
            Bailando~\cite{siyao2022bailando}  & 4.78 & 5.22&5.11&5.11 &5.22 \\
            EDGE~\cite{tseng2023edge}  & 5.67 & 5.00&5.56&5.78 &5.33 \\
            FineNet~\cite{li2023finedance} & 6.44 & 5.89&6.11& 5.33&6.56 \\
            \textbf{SoulNet} & \underline{7.33} & \underline{8.45} & \underline{7.56} &\underline{\textbf{7.67}}&\underline{7.78} \\
            \bottomrule [1pt] 
        \end{tabular}}
        \caption{\textbf{Results of the User Study.} The top of the table presents user evaluation results for different datasets; The bottom presents the evaluation results for dance generation using different methods on \textit{SoulDance} dataset.}
    \label{tab:user_study} 
    \vspace{-1.5em}
\end{table}




\section{Conclusion}
In this paper, we introduce \textit{SoulDance}, a large-scale, high-quality 3D holistic dance dataset for music-driven dance generation, capturing synchronized body movements, hand gestures, and facial expressions. We further present \textit{SoulNet}, which utilizes HRVQ for joint modeling and efficient quantization of holistic dance. Using the music-dance alignment prior from MMR to supervise MAGM, \textit{SoulNet} generates expressive holistic dance sequences that are aligned with the input music. Additionally, we propose two novel evaluation metrics: EmotionAlign and the MMR-Matching Score, to assess the alignment of facial and body motions with music. Both quantitative and qualitative results demonstrate that \textit{SoulNet} achieves SOTA performance in generating holistic dance aligned with music. 

\section*{Acknowledgements}
We sincerely thank the anonymous ICCV reviewers for their invaluable feedback and suggestions. Xiaojie Li would also like to thank his colleagues in the lab, Weixiang Zhang, Rongwei Lu, Jiajun Luo and Duo Wu, for their constructive comments on improving the quality of this paper. This work was supported in part by National Key Research and Development Project of China (Grant No. 2023YFF0905502), National Natural Science Foundation of China (Grant No. 92467204, 62472249), National Key R\&D Program of China (No. 2022YFF0902303), and Shenzhen Science and Technology Program (Grant No. JCYJ20220818101014030 and KJZD20240903102300001).

{   
    \small
    \bibliographystyle{ieeenat_fullname}
    \bibliography{main}

\begin{thebibliography}{62}
\providecommand{\natexlab}[1]{#1}
\providecommand{\url}[1]{\texttt{#1}}
\expandafter\ifx\csname urlstyle\endcsname\relax
  \providecommand{\doi}[1]{doi: #1}\else
  \providecommand{\doi}{doi: \begingroup \urlstyle{rm}\Url}\fi

\bibitem[Arikan and Forsyth(2002)]{arikan2002interactive}
Okan Arikan and David~A Forsyth.
\newblock Interactive motion generation from examples.
\newblock \emph{ACM Transactions on Graphics (TOG)}, 21\penalty0 (3):\penalty0 483--490, 2002.

\bibitem[Au et~al.(2022)Au, Chen, Jiang, and Guo]{au2022choreograph}
Ho~Yin Au, Jie Chen, Junkun Jiang, and Yike Guo.
\newblock Choreograph: Music-conditioned automatic dance choreography over a style and tempo consistent dynamic graph.
\newblock In \emph{Proceedings of the 30th ACM International Conference on Multimedia}, pages 3917--3925, 2022.

\bibitem[{Autodesk, Inc.}(2023)]{motionbuilder}
{Autodesk, Inc.}
\newblock Autodesk motionbuilder, 2023.
\newblock Available at \url{https://www.autodesk.com/products/motionbuilder/overview}.

\bibitem[Barnes et~al.(1996)Barnes, Rizvi, and Nasrabadi]{barnes1996advances}
Christopher~F Barnes, Syed~A Rizvi, and Nasser~M Nasrabadi.
\newblock Advances in residual vector quantization: A review.
\newblock \emph{IEEE transactions on image processing}, 5\penalty0 (2):\penalty0 226--262, 1996.

\bibitem[Baruch et~al.(2021)Baruch, Chen, Dehghan, Dimry, Feigin, Fu, Gebauer, Joffe, Kurz, Schwartz, et~al.]{baruch2021arkitscenes}
Gilad Baruch, Zhuoyuan Chen, Afshin Dehghan, Tal Dimry, Yuri Feigin, Peter Fu, Thomas Gebauer, Brandon Joffe, Daniel Kurz, Arik Schwartz, et~al.
\newblock Arkitscenes: A diverse real-world dataset for 3d indoor scene understanding using mobile rgb-d data.
\newblock \emph{arXiv preprint arXiv:2111.08897}, 2021.

\bibitem[Borsos et~al.(2023)Borsos, Marinier, Vincent, Kharitonov, Pietquin, Sharifi, Roblek, Teboul, Grangier, Tagliasacchi, et~al.]{borsos2023audiolm}
Zal{\'a}n Borsos, Rapha{\"e}l Marinier, Damien Vincent, Eugene Kharitonov, Olivier Pietquin, Matt Sharifi, Dominik Roblek, Olivier Teboul, David Grangier, Marco Tagliasacchi, et~al.
\newblock Audiolm: a language modeling approach to audio generation.
\newblock \emph{IEEE/ACM Transactions on Audio, Speech, and Language Processing}, 2023.

\bibitem[Chen et~al.(2021)Chen, Tan, Lei, Zhang, Guo, Zhang, and Hu]{chen2021choreomaster}
Kang Chen, Zhipeng Tan, Jin Lei, Song-Hai Zhang, Yuan-Chen Guo, Weidong Zhang, and Shi-Min Hu.
\newblock Choreomaster: choreography-oriented music-driven dance synthesis.
\newblock \emph{ACM Transactions on Graphics (TOG)}, 40\penalty0 (4):\penalty0 1--13, 2021.

\bibitem[Dhariwal et~al.(2020)Dhariwal, Jun, Payne, Kim, Radford, and Sutskever]{dhariwal2020jukebox}
Prafulla Dhariwal, Heewoo Jun, Christine Payne, Jong~Wook Kim, Alec Radford, and Ilya Sutskever.
\newblock Jukebox: A generative model for music.
\newblock \emph{arXiv preprint arXiv:2005.00341}, 2020.

\bibitem[Ekman and Friesen(1971)]{ekman1971constants}
Paul Ekman and Wallace~V Friesen.
\newblock Constants across cultures in the face and emotion.
\newblock \emph{Journal of personality and social psychology}, 17\penalty0 (2):\penalty0 124, 1971.

\bibitem[{Epic Games}(2022)]{unrealengine5}
{Epic Games}.
\newblock Unreal engine 5, 2022.
\newblock Available at \url{https://www.unrealengine.com/}.

\bibitem[Fan et~al.(2022)Fan, Lin, Saito, Wang, and Komura]{fan2022faceformer}
Yingruo Fan, Zhaojiang Lin, Jun Saito, Wenping Wang, and Taku Komura.
\newblock Faceformer: Speech-driven 3d facial animation with transformers.
\newblock In \emph{Proceedings of the IEEE/CVF Conference on Computer Vision and Pattern Recognition}, pages 18770--18780, 2022.

\bibitem[Fang et~al.(2017)Fang, Xie, Tai, and Lu]{fang2017rmpe}
Hao-Shu Fang, Shuqin Xie, Yu-Wing Tai, and Cewu Lu.
\newblock Rmpe: Regional multi-person pose estimation.
\newblock In \emph{Proceedings of the IEEE international conference on computer vision}, pages 2334--2343, 2017.

\bibitem[Fink et~al.(2021)Fink, Bl{\"a}sing, Ravignani, and Shackelford]{fink2021evolution}
Bernhard Fink, Bettina Bl{\"a}sing, Andrea Ravignani, and Todd~K Shackelford.
\newblock Evolution and functions of human dance.
\newblock \emph{Evolution and Human Behavior}, 42\penalty0 (4):\penalty0 351--360, 2021.

\bibitem[Gao et~al.(2023)Gao, Hu, Zhang, Zhang, and Bo]{gao2023dancemeld}
Xin Gao, Li Hu, Peng Zhang, Bang Zhang, and Liefeng Bo.
\newblock Dancemeld: Unraveling dance phrases with hierarchical latent codes for music-to-dance synthesis.
\newblock \emph{arXiv preprint arXiv:2401.10242}, 2023.

\bibitem[Gray and Neuhoff(1998)]{gray1998quantization}
Robert~M. Gray and David~L. Neuhoff.
\newblock Quantization.
\newblock \emph{IEEE transactions on information theory}, 44, 1998.

\bibitem[Guo et~al.(2020)Guo, Zuo, Wang, Zou, Sun, Deng, Gong, and Cheng]{guo2020action2motion}
Chuan Guo, Xinxin Zuo, Sen Wang, Shihao Zou, Qingyao Sun, Annan Deng, Minglun Gong, and Li Cheng.
\newblock Action2motion: Conditioned generation of 3d human motions.
\newblock In \emph{Proceedings of the 28th ACM International Conference on Multimedia}, pages 2021--2029, 2020.

\bibitem[Guo et~al.(2022{\natexlab{a}})Guo, Zou, Zuo, Wang, Ji, Li, and Cheng]{guo2022generating}
Chuan Guo, Shihao Zou, Xinxin Zuo, Sen Wang, Wei Ji, Xingyu Li, and Li Cheng.
\newblock Generating diverse and natural 3d human motions from text.
\newblock In \emph{Proceedings of the IEEE/CVF Conference on Computer Vision and Pattern Recognition}, pages 5152--5161, 2022{\natexlab{a}}.

\bibitem[Guo et~al.(2022{\natexlab{b}})Guo, Zuo, Wang, and Cheng]{guo2022tm2t}
Chuan Guo, Xinxin Zuo, Sen Wang, and Li Cheng.
\newblock Tm2t: Stochastic and tokenized modeling for the reciprocal generation of 3d human motions and texts.
\newblock In \emph{European Conference on Computer Vision}, pages 580--597. Springer, 2022{\natexlab{b}}.

\bibitem[Guo et~al.(2024)Guo, Mu, Javed, Wang, and Cheng]{guo2024momask}
Chuan Guo, Yuxuan Mu, Muhammad~Gohar Javed, Sen Wang, and Li Cheng.
\newblock Momask: Generative masked modeling of 3d human motions.
\newblock In \emph{Proceedings of the IEEE/CVF Conference on Computer Vision and Pattern Recognition}, pages 1900--1910, 2024.

\bibitem[Huijben et~al.(2024)Huijben, Douze, Muckley, Van~Sloun, and Verbeek]{huijben2024residual}
Iris~AM Huijben, Matthijs Douze, Matthew Muckley, Ruud~JG Van~Sloun, and Jakob Verbeek.
\newblock Residual quantization with implicit neural codebooks.
\newblock \emph{arXiv preprint arXiv:2401.14732}, 2024.

\bibitem[Jang et~al.(2016)Jang, Gu, and Poole]{jang2016categorical}
Eric Jang, Shixiang Gu, and Ben Poole.
\newblock Categorical reparameterization with gumbel-softmax.
\newblock \emph{arXiv preprint arXiv:1611.01144}, 2016.

\bibitem[Jiang et~al.(2023)Jiang, Chen, Liu, Yu, Yu, and Chen]{jiang2023motiongpt}
Biao Jiang, Xin Chen, Wen Liu, Jingyi Yu, Gang Yu, and Tao Chen.
\newblock Motiongpt: Human motion as a foreign language.
\newblock \emph{arXiv preprint arXiv:2306.14795}, 2023.

\bibitem[Jin et~al.(2017)Jin, Zhang, Li, Tian, Zhu, and Fang]{jin2017towards}
Yanghua Jin, Jiakai Zhang, Minjun Li, Yingtao Tian, Huachun Zhu, and Zhihao Fang.
\newblock Towards the automatic anime characters creation with generative adversarial networks.
\newblock \emph{arXiv preprint arXiv:1708.05509}, 2017.

\bibitem[Kico et~al.(2018)Kico, Grammalidis, Christidis, and Liarokapis]{kico2018digitization}
Iris Kico, Nikos Grammalidis, Yiannis Christidis, and Fotis Liarokapis.
\newblock Digitization and visualization of folk dances in cultural heritage: A review.
\newblock \emph{Inventions}, 3\penalty0 (4):\penalty0 72, 2018.

\bibitem[Kim et~al.(2023)Kim, Kim, and Choi]{kim2023flame}
Jihoon Kim, Jiseob Kim, and Sungjoon Choi.
\newblock Flame: Free-form language-based motion synthesis \& editing.
\newblock In \emph{Proceedings of the AAAI Conference on Artificial Intelligence}, pages 8255--8263, 2023.

\bibitem[Kim et~al.(2006)Kim, Fouad, and Hahn]{kim2006making}
Jae~Woo Kim, Hesham Fouad, and James~K Hahn.
\newblock Making them dance.
\newblock In \emph{AAAI Fall Symposium: Aurally Informed Performance}, page~2, 2006.

\bibitem[Kim et~al.(2003)Kim, Park, and Shin]{kim2003rhythmic}
Tae-hoon Kim, Sang~Il Park, and Sung~Yong Shin.
\newblock Rhythmic-motion synthesis based on motion-beat analysis.
\newblock \emph{ACM Transactions on Graphics (TOG)}, 22\penalty0 (3):\penalty0 392--401, 2003.

\bibitem[Klug et~al.(2020)Klug, Einfalt, Brehm, and Lienhart]{klug2020error}
Nikolas Klug, Moritz Einfalt, Stephan Brehm, and Rainer Lienhart.
\newblock Error bounds of projection models in weakly supervised 3d human pose estimation.
\newblock In \emph{2020 International Conference on 3D Vision (3DV)}, pages 898--907. IEEE, 2020.

\bibitem[Le et~al.(2023)Le, Pham, Do, Tjiputra, Tran, and Nguyen]{le2023music}
Nhat Le, Thang Pham, Tuong Do, Erman Tjiputra, Quang~D Tran, and Anh Nguyen.
\newblock Music-driven group choreography.
\newblock In \emph{Proceedings of the IEEE/CVF Conference on Computer Vision and Pattern Recognition}, pages 8673--8682, 2023.

\bibitem[Li et~al.(2022)Li, Zhao, Zhelun, and Sheng]{li2022danceformer}
Buyu Li, Yongchi Zhao, Shi Zhelun, and Lu Sheng.
\newblock Danceformer: Music conditioned 3d dance generation with parametric motion transformer.
\newblock In \emph{Proceedings of the AAAI Conference on Artificial Intelligence}, pages 1272--1279, 2022.

\bibitem[Li et~al.(2021)Li, Yang, Ross, and Kanazawa]{li2021ai}
Ruilong Li, Shan Yang, David~A Ross, and Angjoo Kanazawa.
\newblock Ai choreographer: Music conditioned 3d dance generation with aist++.
\newblock In \emph{Proceedings of the IEEE/CVF International Conference on Computer Vision}, pages 13401--13412, 2021.

\bibitem[Li et~al.(2023)Li, Zhao, Zhang, Su, Ren, Zhang, Tang, and Li]{li2023finedance}
Ronghui Li, Junfan Zhao, Yachao Zhang, Mingyang Su, Zeping Ren, Han Zhang, Yansong Tang, and Xiu Li.
\newblock Finedance: A fine-grained choreography dataset for 3d full body dance generation.
\newblock In \emph{Proceedings of the IEEE/CVF International Conference on Computer Vision}, pages 10234--10243, 2023.

\bibitem[Li et~al.(2024)Li, Zhang, Zhang, Zhang, Guo, Zhang, Liu, and Li]{li2024lodge}
Ronghui Li, YuXiang Zhang, Yachao Zhang, Hongwen Zhang, Jie Guo, Yan Zhang, Yebin Liu, and Xiu Li.
\newblock Lodge: A coarse to fine diffusion network for long dance generation guided by the characteristic dance primitives.
\newblock In \emph{Proceedings of the IEEE/CVF Conference on Computer Vision and Pattern Recognition}, pages 1524--1534, 2024.

\bibitem[Li et~al.(2017{\natexlab{a}})Li, Bolkart, Black, Li, and Romero]{FLAME:SiggraphAsia2017}
Tianye Li, Timo Bolkart, Michael.~J. Black, Hao Li, and Javier Romero.
\newblock Learning a model of facial shape and expression from {4D} scans.
\newblock \emph{ACM Transactions on Graphics, (Proc. SIGGRAPH Asia)}, 36\penalty0 (6), 2017{\natexlab{a}}.

\bibitem[Li et~al.(2017{\natexlab{b}})Li, Bolkart, Black, Li, and Romero]{li2017learning}
Tianye Li, Timo Bolkart, Michael~J Black, Hao Li, and Javier Romero.
\newblock Learning a model of facial shape and expression from 4d scans.
\newblock \emph{ACM Trans. Graph.}, 36\penalty0 (6):\penalty0 194--1, 2017{\natexlab{b}}.

\bibitem[Liu et~al.(2024)Liu, Zhu, Becherini, Peng, Su, Zhou, Zhe, Iwamoto, Zheng, and Black]{liu2024emage}
Haiyang Liu, Zihao Zhu, Giorgio Becherini, Yichen Peng, Mingyang Su, You Zhou, Xuefei Zhe, Naoya Iwamoto, Bo Zheng, and Michael~J Black.
\newblock Emage: Towards unified holistic co-speech gesture generation via expressive masked audio gesture modeling.
\newblock In \emph{Proceedings of the IEEE/CVF Conference on Computer Vision and Pattern Recognition}, pages 1144--1154, 2024.

\bibitem[Loper et~al.(2015)Loper, Mahmood, Romero, Pons-Moll, and Black]{SMPL2015}
Matthew Loper, Naureen Mahmood, Javier Romero, Gerard Pons-Moll, and Michael~J. Black.
\newblock {SMPL}: A skinned multi-person linear model.
\newblock \emph{ACM Trans. Graphics (Proc. SIGGRAPH Asia)}, 34\penalty0 (6):\penalty0 248:1--248:16, 2015.

\bibitem[Lu et~al.(2023)Lu, Chen, Zeng, Lin, Zhang, Zhang, and Shum]{lu2023humantomato}
Shunlin Lu, Ling-Hao Chen, Ailing Zeng, Jing Lin, Ruimao Zhang, Lei Zhang, and Heung-Yeung Shum.
\newblock Humantomato: Text-aligned whole-body motion generation.
\newblock \emph{arXiv preprint arXiv:2310.12978}, 2023.

\bibitem[Pavlakos et~al.(2019{\natexlab{a}})Pavlakos, Choutas, Ghorbani, Bolkart, Osman, Tzionas, and Black]{pavlakos2019expressive}
Georgios Pavlakos, Vasileios Choutas, Nima Ghorbani, Timo Bolkart, Ahmed~AA Osman, Dimitrios Tzionas, and Michael~J Black.
\newblock Expressive body capture: 3d hands, face, and body from a single image.
\newblock In \emph{Proceedings of the IEEE/CVF conference on computer vision and pattern recognition}, pages 10975--10985, 2019{\natexlab{a}}.

\bibitem[Pavlakos et~al.(2019{\natexlab{b}})Pavlakos, Choutas, Ghorbani, Bolkart, Osman, Tzionas, and Black]{SMPL-X:2019}
Georgios Pavlakos, Vasileios Choutas, Nima Ghorbani, Timo Bolkart, Ahmed A.~A. Osman, Dimitrios Tzionas, and Michael~J. Black.
\newblock Expressive body capture: 3d hands, face, and body from a single image.
\newblock In \emph{Proceedings IEEE Conf. on Computer Vision and Pattern Recognition (CVPR)}, 2019{\natexlab{b}}.

\bibitem[Petrovich et~al.(2022)Petrovich, Black, and Varol]{petrovich2022temos}
Mathis Petrovich, Michael~J Black, and G{\"u}l Varol.
\newblock Temos: Generating diverse human motions from textual descriptions.
\newblock In \emph{European Conference on Computer Vision}, pages 480--497. Springer, 2022.

\bibitem[Petrovich et~al.(2023)Petrovich, Black, and Varol]{petrovich2023tmr}
Mathis Petrovich, Michael~J Black, and G{\"u}l Varol.
\newblock Tmr: Text-to-motion retrieval using contrastive 3d human motion synthesis.
\newblock In \emph{Proceedings of the IEEE/CVF International Conference on Computer Vision}, pages 9488--9497, 2023.

\bibitem[Radford et~al.(2021)Radford, Kim, Hallacy, Ramesh, Goh, Agarwal, Sastry, Askell, Mishkin, Clark, et~al.]{radford2021learning}
Alec Radford, Jong~Wook Kim, Chris Hallacy, Aditya Ramesh, Gabriel Goh, Sandhini Agarwal, Girish Sastry, Amanda Askell, Pamela Mishkin, Jack Clark, et~al.
\newblock Learning transferable visual models from natural language supervision.
\newblock In \emph{International Conference on Machine Learning}, pages 8748--8763. PMLR, 2021.

\bibitem[Rombach et~al.(2022)Rombach, Blattmann, Lorenz, Esser, and Ommer]{rombach2022high}
Robin Rombach, Andreas Blattmann, Dominik Lorenz, Patrick Esser, and Bj{\"o}rn Ommer.
\newblock High-resolution image synthesis with latent diffusion models.
\newblock In \emph{Proceedings of the IEEE/CVF Conference on Computer Vision and Pattern Recognition}, pages 10684--10695, 2022.

\bibitem[Saharia et~al.(2022)Saharia, Chan, Saxena, Li, Whang, Denton, Ghasemipour, Gontijo~Lopes, Karagol~Ayan, Salimans, et~al.]{saharia2022photorealistic}
Chitwan Saharia, William Chan, Saurabh Saxena, Lala Li, Jay Whang, Emily~L Denton, Kamyar Ghasemipour, Raphael Gontijo~Lopes, Burcu Karagol~Ayan, Tim Salimans, et~al.
\newblock Photorealistic text-to-image diffusion models with deep language understanding.
\newblock \emph{Advances in neural information processing systems}, 35:\penalty0 36479--36494, 2022.

\bibitem[Savchenko(2023)]{savchenko2023facial}
Andrey Savchenko.
\newblock Facial expression recognition with adaptive frame rate based on multiple testing correction.
\newblock In \emph{Proceedings of the 40th International Conference on Machine Learning (ICML)}, pages 30119--30129. PMLR, 2023.

\bibitem[{SHANGHAI CHINGMU VISION TECHNOLOGY}(2022)]{chingmu}
{SHANGHAI CHINGMU VISION TECHNOLOGY}.
\newblock Chingmu, 2022.
\newblock Available at \url{https://www.chingmu.com/}.

\bibitem[Shiratori et~al.(2006)Shiratori, Nakazawa, and Ikeuchi]{shiratori2006dancing}
Takaaki Shiratori, Atsushi Nakazawa, and Katsushi Ikeuchi.
\newblock Dancing-to-music character animation.
\newblock In \emph{Computer Graphics Forum}, pages 449--458. Wiley Online Library, 2006.

\bibitem[Siyao et~al.(2022)Siyao, Yu, Gu, Lin, Wang, Qian, Loy, and Liu]{siyao2022bailando}
Li Siyao, Weijiang Yu, Tianpei Gu, Chunze Lin, Quan Wang, Chen Qian, Chen~Change Loy, and Ziwei Liu.
\newblock Bailando: 3d dance generation by actor-critic gpt with choreographic memory.
\newblock In \emph{Proceedings of the IEEE/CVF Conference on Computer Vision and Pattern Recognition}, pages 11050--11059, 2022.

\bibitem[Sun et~al.(2022)Sun, Cao, Jiang, Yuan, Bai, Kitani, and Luo]{sun2022dancetrack}
Peize Sun, Jinkun Cao, Yi Jiang, Zehuan Yuan, Song Bai, Kris Kitani, and Ping Luo.
\newblock Dancetrack: Multi-object tracking in uniform appearance and diverse motion.
\newblock In \emph{Proceedings of the IEEE/CVF Conference on Computer Vision and Pattern Recognition}, pages 20993--21002, 2022.

\bibitem[Tang et~al.(2018)Tang, Jia, and Mao]{tang2018dance}
Taoran Tang, Jia Jia, and Hanyang Mao.
\newblock Dance with melody: An lstm-autoencoder approach to music-oriented dance synthesis.
\newblock In \emph{Proceedings of the 26th ACM international conference on Multimedia}, pages 1598--1606, 2018.

\bibitem[Tseng et~al.(2023)Tseng, Castellon, and Liu]{tseng2023edge}
Jonathan Tseng, Rodrigo Castellon, and Karen Liu.
\newblock Edge: Editable dance generation from music.
\newblock In \emph{Proceedings of the IEEE/CVF Conference on Computer Vision and Pattern Recognition}, pages 448--458, 2023.

\bibitem[Valle-P{\'e}rez et~al.(2021)Valle-P{\'e}rez, Henter, Beskow, Holzapfel, Oudeyer, and Alexanderson]{valle2021transflower}
Guillermo Valle-P{\'e}rez, Gustav~Eje Henter, Jonas Beskow, Andre Holzapfel, Pierre-Yves Oudeyer, and Simon Alexanderson.
\newblock Transflower: probabilistic autoregressive dance generation with multimodal attention.
\newblock \emph{ACM Transactions on Graphics (TOG)}, 40\penalty0 (6):\penalty0 1--14, 2021.

\bibitem[Van Den~Oord et~al.(2017)Van Den~Oord, Vinyals, et~al.]{van2017neural}
Aaron Van Den~Oord, Oriol Vinyals, et~al.
\newblock Neural discrete representation learning.
\newblock \emph{Advances in Neural Information Processing Systems}, 30, 2017.

\bibitem[Xing et~al.(2023)Xing, Xia, Zhang, Cun, Wang, and Wong]{xing2023codetalker}
Jinbo Xing, Menghan Xia, Yuechen Zhang, Xiaodong Cun, Jue Wang, and Tien-Tsin Wong.
\newblock Codetalker: Speech-driven 3d facial animation with discrete motion prior.
\newblock In \emph{Proceedings of the IEEE/CVF Conference on Computer Vision and Pattern Recognition}, pages 12780--12790, 2023.

\bibitem[Xue et~al.(2022)Xue, Wang, Tan, Ma, and Guo]{xue2022vision}
Fanglei Xue, Qiangchang Wang, Zichang Tan, Zhongsong Ma, and Guodong Guo.
\newblock Vision transformer with attentive pooling for robust facial expression recognition.
\newblock \emph{IEEE Transactions on Affective Computing}, 2022.

\bibitem[Yao et~al.(2024)Yao, Song, Zhou, Ao, Chen, and Liu]{yao2024moconvq}
Heyuan Yao, Zhenhua Song, Yuyang Zhou, Tenglong Ao, Baoquan Chen, and Libin Liu.
\newblock Moconvq: Unified physics-based motion control via scalable discrete representations.
\newblock \emph{ACM Transactions on Graphics (TOG)}, 43\penalty0 (4):\penalty0 1--21, 2024.

\bibitem[Zeghidour et~al.(2021)Zeghidour, Luebs, Omran, Skoglund, and Tagliasacchi]{zeghidour2021soundstream}
Neil Zeghidour, Alejandro Luebs, Ahmed Omran, Jan Skoglund, and Marco Tagliasacchi.
\newblock Soundstream: An end-to-end neural audio codec.
\newblock \emph{IEEE/ACM Transactions on Audio, Speech, and Language Processing}, 30:\penalty0 495--507, 2021.

\bibitem[Zhang et~al.(2024)Zhang, Tang, Zhang, Lin, Han, Xiao, and Wang]{zhang2024bidirectional}
Canyu Zhang, Youbao Tang, Ning Zhang, Ruei-Sung Lin, Mei Han, Jing Xiao, and Song Wang.
\newblock Bidirectional autoregessive diffusion model for dance generation.
\newblock In \emph{Proceedings of the IEEE/CVF Conference on Computer Vision and Pattern Recognition}, pages 687--696, 2024.

\bibitem[Zhang et~al.(2023)Zhang, Zhang, Cun, Huang, Zhang, Zhao, Lu, and Shen]{zhang2023t2m}
Jianrong Zhang, Yangsong Zhang, Xiaodong Cun, Shaoli Huang, Yong Zhang, Hongwei Zhao, Hongtao Lu, and Xi Shen.
\newblock T2m-gpt: Generating human motion from textual descriptions with discrete representations.
\newblock \emph{arXiv preprint arXiv:2301.06052}, 2023.

\bibitem[Zhou et~al.(2019)Zhou, Barnes, Lu, Yang, and Li]{zhou2019continuity}
Yi Zhou, Connelly Barnes, Jingwan Lu, Jimei Yang, and Hao Li.
\newblock On the continuity of rotation representations in neural networks.
\newblock In \emph{Proceedings of the IEEE/CVF conference on computer vision and pattern recognition}, pages 5745--5753, 2019.

\bibitem[Zhuang et~al.(2022)Zhuang, Wang, Chai, Wang, Shao, and Xia]{zhuang2022music2dance}
Wenlin Zhuang, Congyi Wang, Jinxiang Chai, Yangang Wang, Ming Shao, and Siyu Xia.
\newblock Music2dance: Dancenet for music-driven dance generation.
\newblock \emph{ACM Transactions on Multimedia Computing, Communications, and Applications (TOMM)}, 18\penalty0 (2):\penalty0 1--21, 2022.

\end{thebibliography}
}

\clearpage
\appendix

\setcounter{page}{1}

\twocolumn[
\centering
\Large
\textbf{Music-Aligned Holistic 3D Dance Generation via Hierarchical Motion Modeling} \\
\vspace{0.5em}Supplementary Material \\
\vspace{1.0em}]

\appendix

\section{Additional Details of SoulDance Dataset}
\label{supl:add_dataset}
\begin{figure}[th]
	\centering
	\includegraphics[width=\linewidth]{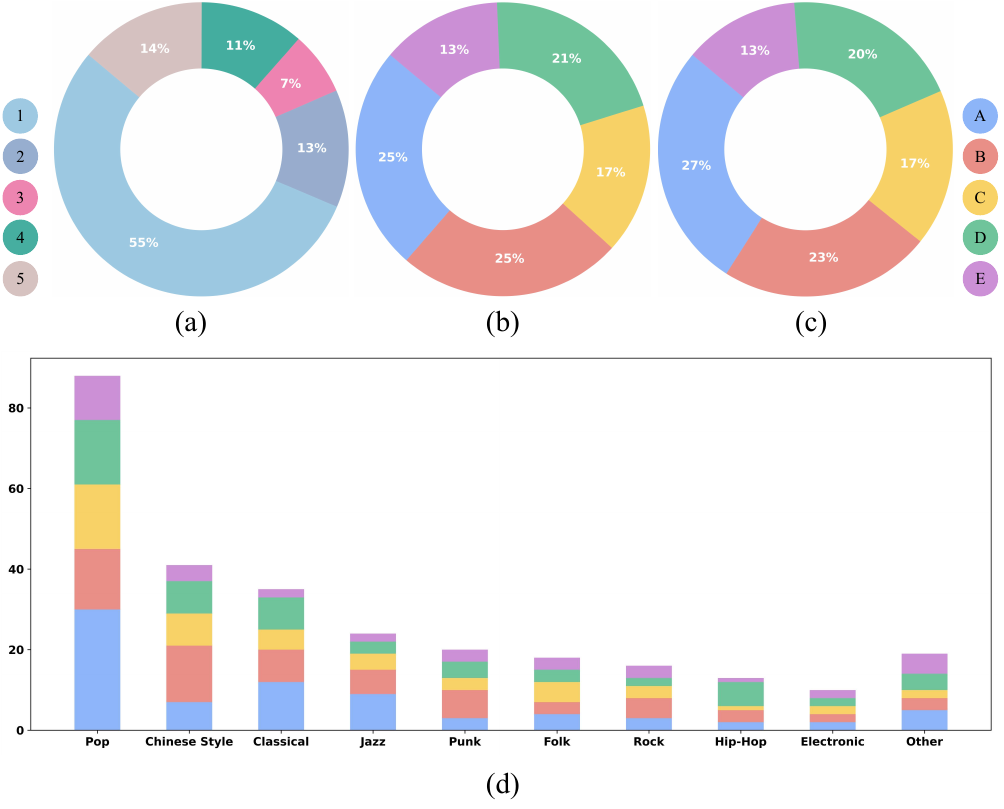}
	\caption{\textbf{Overview of the distribution of the \textit{SoulDance} dataset.} (a) shows the distribution of dance sequences by dancer count (1–5), with most sequences featuring solo performances. (b) depicts the proportion of dance duration for each dancer (A-E). (c) illustrates the distribution of dance sequence counts per dancer. (d) displays the number of dance genres and the count of dance sequences per genre for each dancer.} 
        \label{fig:circle_bar}
\end{figure}

\begin{figure*}[th]
	\centering
	\includegraphics[width=1\linewidth]{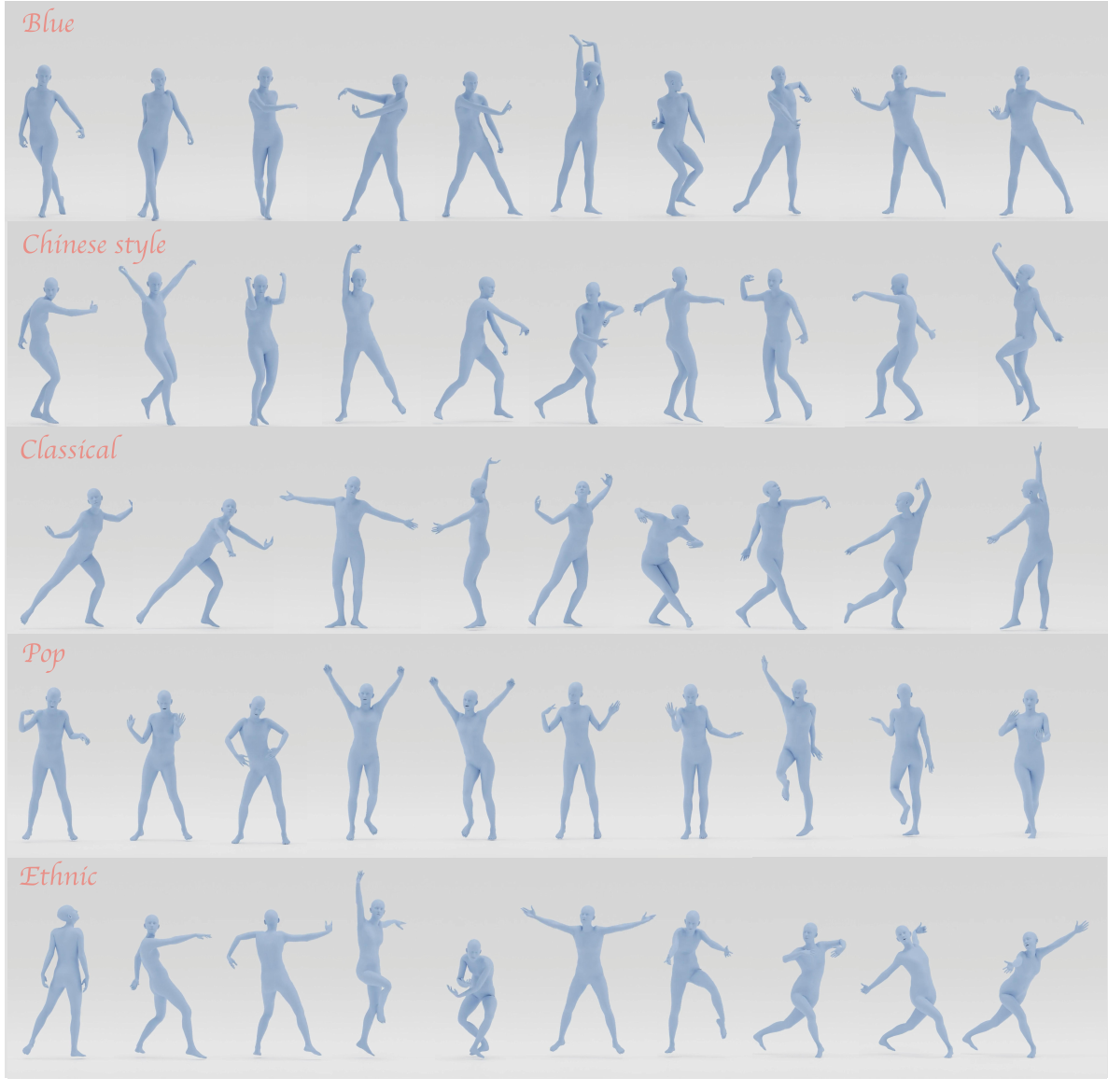}
	\caption{\textbf{Showcase of various dance styles in the \textit{SoulDance} dataset.} The \textit{SoulDance} dataset demonstrates high motion quality and diversity across multiple dance styles.}
        \label{fig:dataset_show}
\end{figure*}

In Figure~\ref{fig:circle_bar}, we illustrate the relationships among dance genres, the number of dancers, and each dancer's proportion within the dataset.  
Figure~\ref{fig:dataset_show} showcases various music-dance motions from different styles in the \textit{SoulDance} dataset. The body and hand movements demonstrate remarkable diversity and precision, further enhanced by expressive facial motions, making the dances more dynamic and emotionally engaging.

\section{Body-Hands Motion Refinement} 
\label{supl:refine}
The majority of existing datasets focus on body movement, typically utilizing 24 body joints from the SMPL model~\cite{SMPL2015}. In contrast, our dataset captures detailed holistic dance motion, requiring the use of the SMPL-X model~\cite{pavlakos2019expressive}, which includes 22 body joints, 30 hand joints and is compatible with FLAME~\cite{li2017learning} parameters for facial expressions. 
As shown in Figure~\ref{fig:teaser_image}, once the body movement and hand gesture BVH data is acquired, we employ MotionBuilder~\cite{motionbuilder} and Unreal Engine 5~\cite{unrealengine5} to retarget motions to the SMPL-X format. Throughout the retargeting process, we implement a workflow within Unreal Engine 5, including T-pose adjustments, bone length calibration, and joint name mapping to ensure precise alignment. When necessary, our team of engineers performs manual refinements to correct any non-physical joint behaviors, further enhancing the authenticity of the retargeted motions.

\section{Transforming Face Blendshapes to FLAME}
\label{supl:flame}
We follow the method introduced in EMAGE~\cite{liu2024emage} to convert ARKit blendshape weights into FLAME parameters. Given the ARKit blendshape weights $\mathbf{b}_{\text{ARKit}} \in \mathbb{R}^{T \times 52}$, we aim to derive a transformation matrix $\mathbf{W} \in \mathbb{R}^{52 \times 103}$ to map these into FLAME parameters $\mathbf{b}_{\text{FLAME}} \in \mathbb{R}^{T \times (100 + 3)}$, where the dimensionality of 100 corresponds to expression parameters, and 3 represents jaw movements. We leverage a set of handcrafted blendshape templates $\mathbf{v}_t \in \mathbb{R}^{52}$ on the FLAME model, structured according to ARKit's Facial Action Coding System (FACS) configuration. This setup enables direct control of the FLAME topology vertices $\mathbf{v}$ using the blendshape weights:
\begin{equation}
\mathbf{v} = \mathbf{v}_t^0 + \sum_{j=1}^{52} \mathbf{b}_{\text{ARKit}, j} \cdot \mathbf{v}_t^j,
\end{equation}
where $\mathbf{b}_{\text{ARKit}, j}$ is the weight of the $j$-th ARKit blendshape, and $\mathbf{v}_t^j$ is the FLAME template vertex position. The term $\mathbf{v}_t^0$ denotes the initial template vertex positions in the FLAME model. We optimize $\mathbf{W}$ by minimizing the Euclidean distance $\| \Tilde{\mathbf{v}}_j - \mathbf{v}_j \|_2$, where $\Tilde{\mathbf{v}}$ represents vertices derived from FLAME's Linear Blend Skinning (LBS) function $\mathcal{V}(\mathbf{b}_{\text{FLAME}})$.

\section{Holistic Dance Motion Representation}
\label{supl:motion_rep}
Following the HumanML3D format~\cite{guo2022generating} and HumanTomato format~\cite{lu2023humantomato} for motion representation, we represent the holistic motion at each frame $m_i$ as a tuple containing various motion attributes. Specifically, we define $m_i$ by the root angular velocity $\dot{r}^a \in \mathbb{R}$ along the Y-axis, root linear velocities $\dot{r}^x, \dot{r}^z \in \mathbb{R}$ on the XZ-plane, root height $r^y \in \mathbb{R}$, local joint positions $\mathbf{j}^p \in \mathbb{R}^{3N-3}$, 6-DOF joint rotations~\cite{zhou2019continuity} $\mathbf{j}^r \in \mathbb{R}^{6N-6}$, joint velocities $\mathbf{j}^v \in \mathbb{R}^{3N}$, and foot contact indicators $\dot{c} \in \mathbb{R}^4$. Here, $N = 52$ represents the total number of body-hand joints, utilizing 22 body joints and 30 hand joints as defined in the SMPL-X model~\cite{pavlakos2019expressive}. For facial motion, we adopt the FLAME format~\cite{kim2023flame}, using $\mathbf{f} \in \mathbb{R}^{100}$ to represent facial expressions. Thus, each frame’s whole-body motion is represented as $\mathbf{m}_i = \{\dot{r}^a, \dot{r}^x, \dot{r}^z, r^y, \mathbf{j}^p, \mathbf{j}^r, \mathbf{j}^v, \dot{c}, \mathbf{f}\}$, with a total dimension of 723.

\section{Dance Reconstruction Evaluation Metrics}
\label{supl:rec_eval} During HRVQ  training, it is essential to evaluate the reconstruction quality of dance. While body and hand movements can be assessed using the standard MPJPE~\cite{klug2020error}, it fails to capture the accuracy of facial reconstruction. To address this, we introduce the Face Vertex Error (FVE), which quantifies the deviation of reconstructed facial sequences from the ground truth~\cite{xing2023codetalker, fan2022faceformer}. FVE is computed by measuring the Euclidean distance between the ground truth and reconstructed facial vertices for each frame, then averaging these distances over the entire sequence: 
\begin{equation}
\text{\textit{FVE}} = \frac{1}{N} \sum_{i=1}^{N} \sqrt{\sum_{j=1}^{V} (v_j - \tilde{v}_j )^2}
\end{equation}
where $ v_j $ represents the ground truth positions of the facial vertices, and $ \tilde{v}_j $ denotes the corresponding vertices reconstructed by HRVQ. The metric is averaged over $ N $ frames to evaluate facial reconstruction quality.

\section{Additional Qualitative Results}
\label{supl:add_vis_res}

\begin{figure*}[th]
	\centering
	\includegraphics[width=0.8\linewidth]{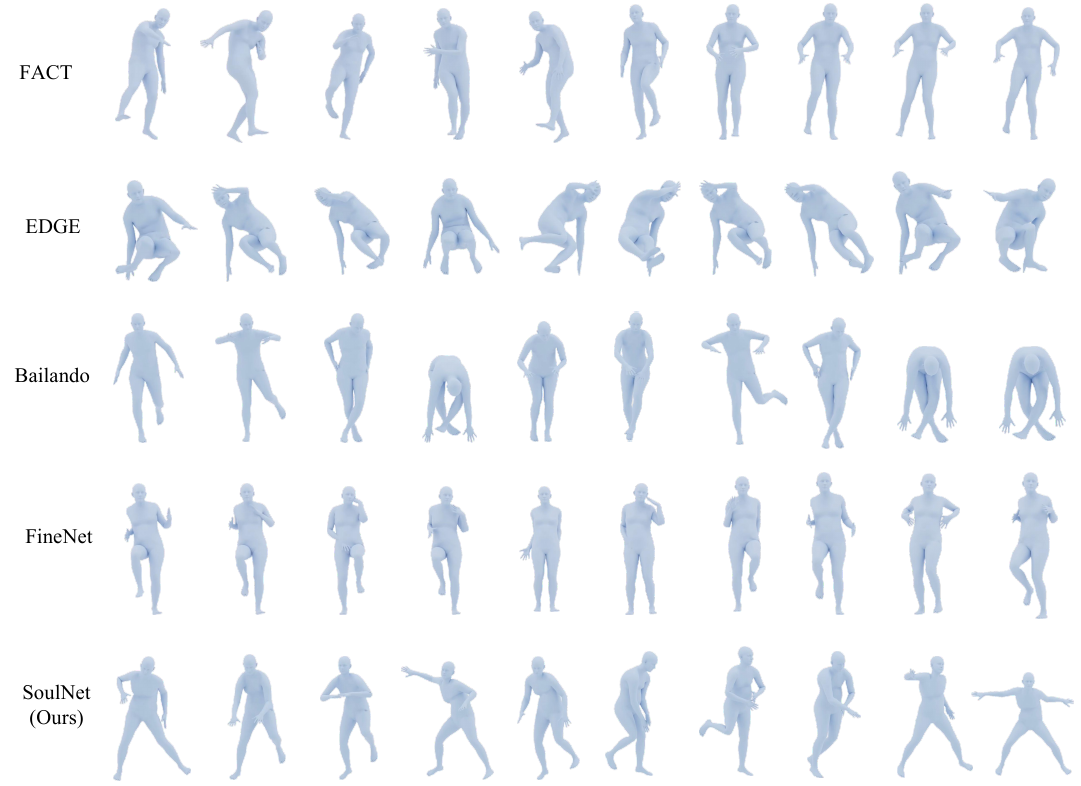}
	\caption{\textbf{Qualitative generation result comparisons} for a \textit{Rock} song in the AIST++ dataset.}
        \label{fig:aistpp_results}
\end{figure*}

\begin{figure*}[th]
	\centering
	\includegraphics[width=0.8\linewidth]{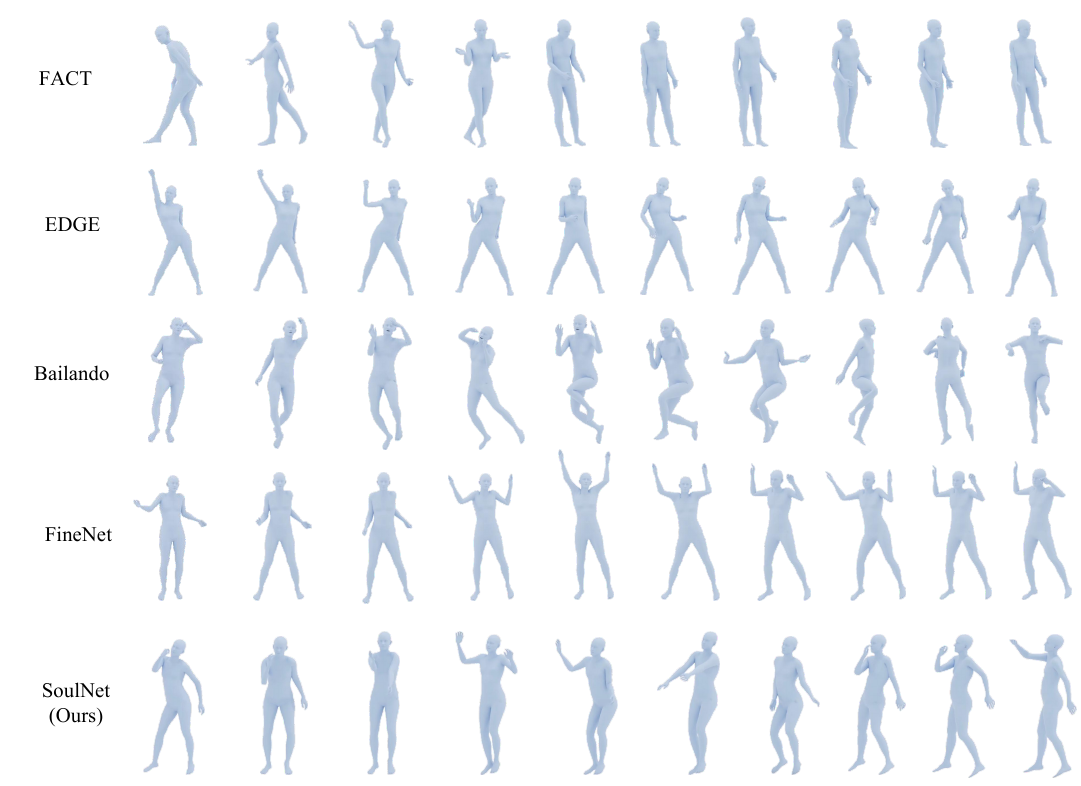}
	\caption{\textbf{Qualitative generation result comparisons} for a \textit{Pop} song in the \textit{SoulDance} dataset.}
        \label{fig:souldance_results}
\end{figure*}

\noindent\textbf{Comparison with SOTA Methods.} \textit{SoulNet} demonstrates exceptional qualitative performance on both the AIST++ and \textit{SoulDance} datasets. In Figure~\ref{fig:souldance_results}, FACT~\cite{li2021ai} generates dance sequences where, after the initial two seconds, body and hand movements become mostly static, and facial expressions are entirely absent. EDGE~\cite{tseng2023edge} produces convincing body movements but often fails to generate detailed hand motions and lacks expressive facial output. Bailando~\cite{siyao2022bailando} captures body movements and facial expressions effectively but suffers from joint dislocations during turns and inadequate hand generation. FineNet~\cite{li2023finedance} delivers satisfactory dance and hand motions but struggles with fine hand articulation and facial expressiveness. 
In contrast, \textit{SoulNet} not only generates diverse and dynamic body movements but also excels in capturing intricate hand and facial details. As shown in Figure~\ref{fig:aistpp_results}, on the AIST++ dataset, SoulNet achieves superior alignment with the musical beat and demonstrates greater diversity in generated dance sequences compared to other methods.

\begin{figure*}[th]
	\centering
	\includegraphics[width=0.8\linewidth]{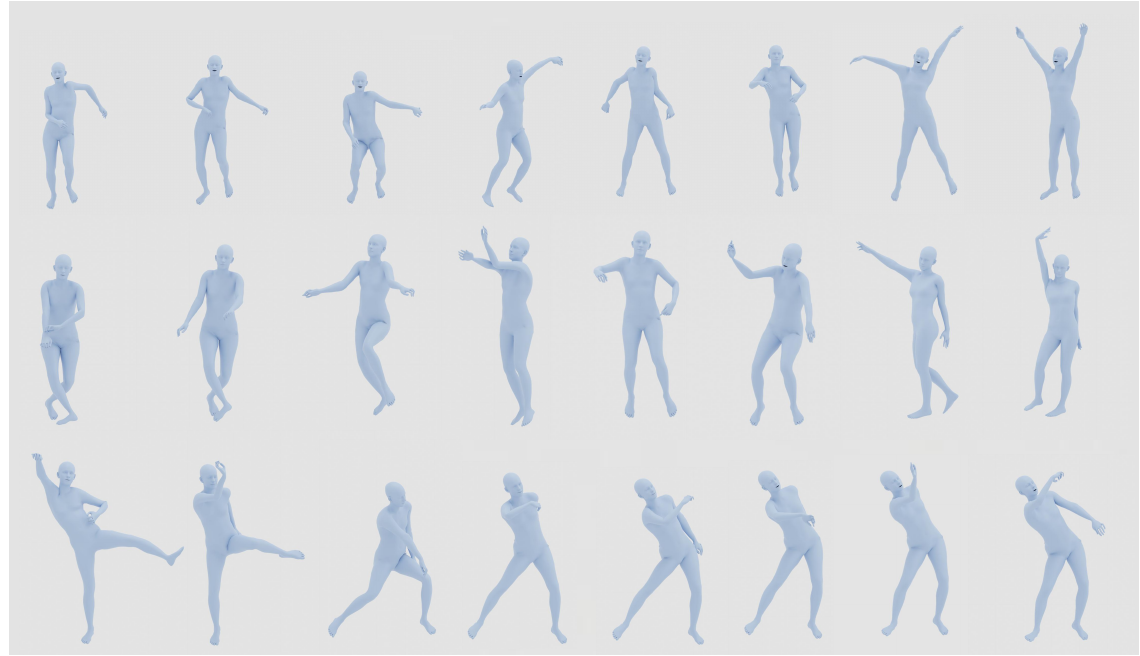}
	\caption{\textbf{Diversity of generated dances.} The \textit{SoulNet} method demonstrates rich diversity under identical input music of the \textit{Chinese Style} genre, encompassing variations in body movements, hand gestures, and facial expressions.}
        \label{fig:soulnet_div}
\end{figure*}

\noindent\textbf{Generating Diverse Dances.} SoulNet is used to generate three dance fragments from the same music clip. As illustrated in Figure~\ref{fig:soulnet_div}, the generated dances exhibit significant diversity and richness in movement while maintaining alignment with the input music genre, showcasing the excellent multimodal capabilities of our method.  

\begin{figure}[h]
	\centering
	\includegraphics[width=1\linewidth]{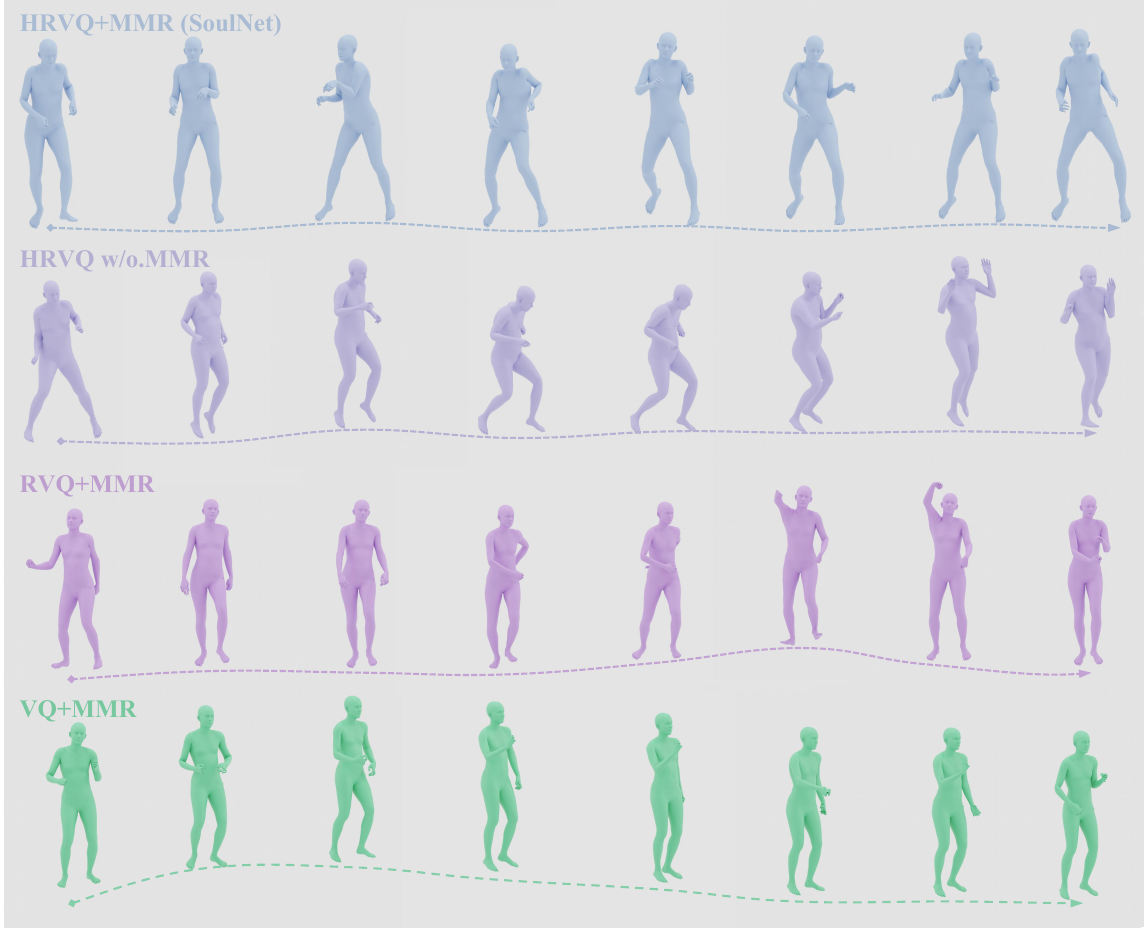}
	\caption{\textbf{Comparison of Visualization Results.} We visualize the dance generation results on the SoulDance dataset using different methods. The dashed line represents the motion trajectory along the direction of gravity, where smaller fluctuations indicate more stable generated dance motions.}
        \label{fig:soulnet_abl}
        \vspace{-1em}
\end{figure}

\begin{figure*}[t]
	\centering
	\includegraphics[width=1\linewidth]{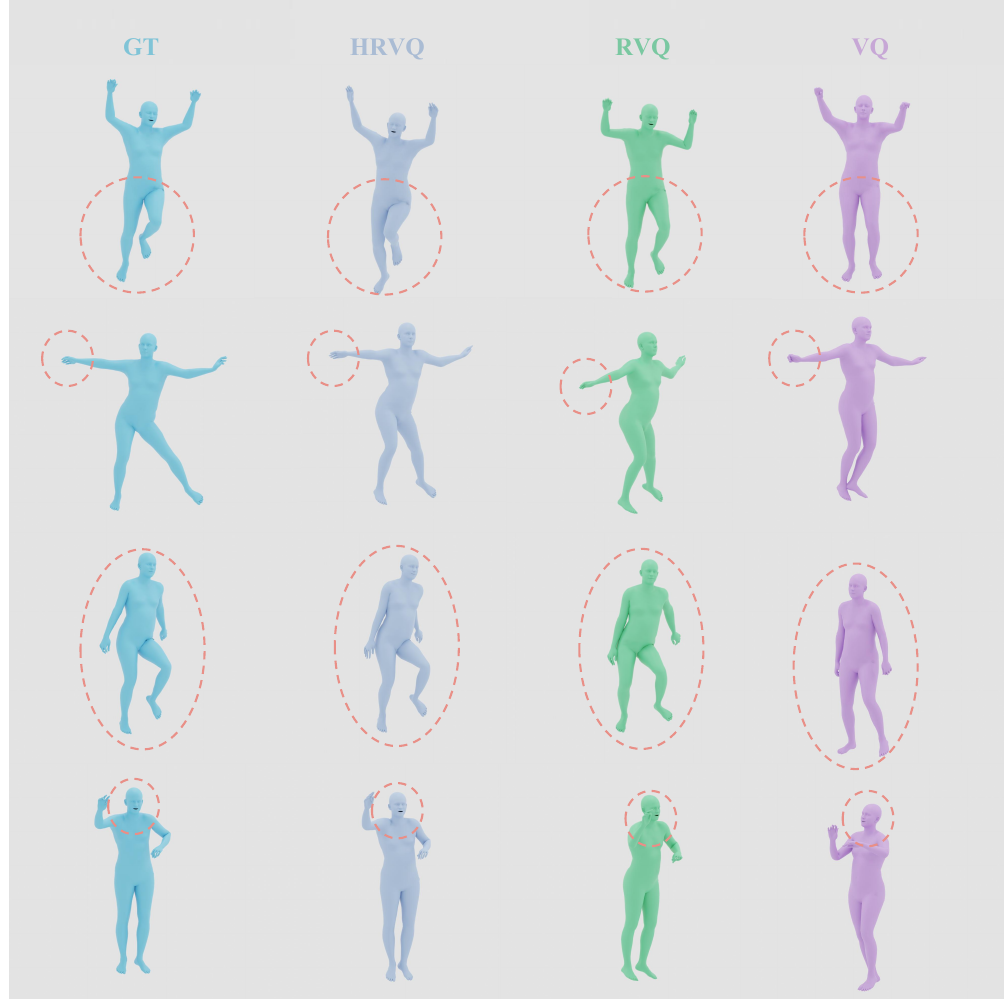}
	\caption{\textbf{Visualizes dance motion reconstruction on the \textit{SoulDance} dataset.} From left to right, the columns represent the ground truth (GT), HRVQ, RVQ, and VQ results, respectively.}
        \label{fig:vis_vqs}
\end{figure*}

\noindent\textbf{Comparison of Different VQ Methods.} Figure~\ref{fig:vis_vqs} presents a comparison of dance motion reconstruction for a ground truth sequence using three quantization methods—HRVQ, RVQ, and VQ—all configured with a codebook size of 512. The results clearly demonstrate that HRVQ outperforms the other methods across all key aspects, including body reconstruction (rows 1 and 3), hand gestures (row 2), and facial expressions (row 4). Furthermore, Figure~\ref{fig:soulnet_abl} shows that HRVQ produces the most stable and consistent dance sequences, followed by RVQ, with VQ performing the worst. These findings underscore HRVQ’s superior ability to capture fine-grained and expressive dance motions, significantly surpassing RVQ and VQ in reconstruction quality.


\section{Implementation Details.}
\label{supl:soulnet_impl}
For HRVQ, we employ residual blocks for both the encoder and decoder, with a downscale factor of 4. Each vector quantizer consists of 6 layers, with each layer’s codebook containing 512-dimensional codes. The transformation process uses a MLP and a 1D convolution. The quantization dropout ratio, $q$, is set to 0.2. For MAGM, we use 6 transformer layers and 6 residual transformer layers, with 8 attention heads and a latent dimension of 512. The learning rate reaches $2 \times 10^{-4}$ after 2000 iterations, following a linear warm-up schedule for training all models. The batch size is set to 256 for training HRVQ and 64 for training MAGM. During inference, we apply a classifier-free guidance (CFG) scale of 4 and 5.
For training MMR module, we use the AdamW optimizer with a learning rate of $1 \times 10^{-4}$ and a batch size of 32. The latent dimensionality of the embeddings is set to $d = 256$. We set the temperature $\tau$ to 0.1, and the weight for the InfoNCE loss to 0.1. The threshold for filtering negatives is set to 0.8. All experiments are conducted on 4 NVIDIA V100 GPUs, and the whole process is completed within three days.

\section{Training: Music-Motion Retrieval Module}
\label{supl:traing_mmr}

\noindent\textbf{Dataset.} To establish robust dance-music alignment priors, we gathered high-quality open-source music-dance datasets: AIST++~\cite{li2021ai}, Finedance~\cite{li2023finedance}, PhantomDance~\cite{li2022danceformer}, and \textit{SoulDance} (totaling 25 hours). Following the preprocessing protocol of EDGE~\cite{tseng2023edge} with default temporal segmentation parameters, we derive two specialized motion representations for training distinct Music-Motion Retrieval modules. For Body-Alignment MMR, all sequences are processed using the HumanML3D~\cite{guo2020action2motion} motion representation ($D_m = 263$). For Whole-Alignment MMR, SoulDance dataset is reformatted via our Holistic Dance Representation (Appendix~\ref{supl:motion_rep}, $D_m = 723$), encoding holistic motion movements. Finally, all datasets are partitioned into training/validation/test splits (8:1:1 ratio) using a stratified strategy that preserves music genre and dance style distributions. 

\noindent\textbf{Technical Details.} Crucially, we enforce temporal alignment constraints between \textit{SoulNet} and MMR training subsets within AIST++ and SoulDance to prevent data leakage—ensuring no overlapping music clips or motion segments exist across models. 
Two specialized MMR modules, $\text{MMR}_{\text{body}}$ and $\text{MMR}_{\text{whole}}$, are pre-trained to provide supervisory signals for the respective losses. $\text{MMR}_{\text{body}}$ is trained on body-only motion data from public datasets (AIST++~\cite{li2021ai}, FineDance~\cite{li2023finedance}], PhantomDance~\cite{li2022danceformer}) using a 263-dimensional motion representation ($D_m = 263$). In contrast, $\text{MMR}_{\text{whole}}$ is trained on a subset of SoulDance holistic motion data (including body, hands and face) with a 723-dimensional representation ($D_m = 723$). 

\noindent\textbf{Training Details.}  
We adopt the same encoder and decoder architectures as TEMOS~\cite{petrovich2022temos} for training our MMR module, with modifications applied only to the encoder dimensions, while keeping the decoder parameters unchanged. Implementation details are consistent with TMR~\cite{petrovich2023tmr}. For optimization, we use the AdamW optimizer with a learning rate of $10^{-4}$ and a batch size of 128, as batch size is a critical hyperparameter for the InfoNCE loss. The latent embedding dimensionality is set to $d = 256$, with the temperature $\tau$ set to 0.1 and the weight of the contrastive loss term $\lambda_{NCE}$ set to 0.1. The threshold for filtering negative samples is configured at 0.8. 

\begin{figure*}[th]
	\centering
	\includegraphics[width=0.8\linewidth]{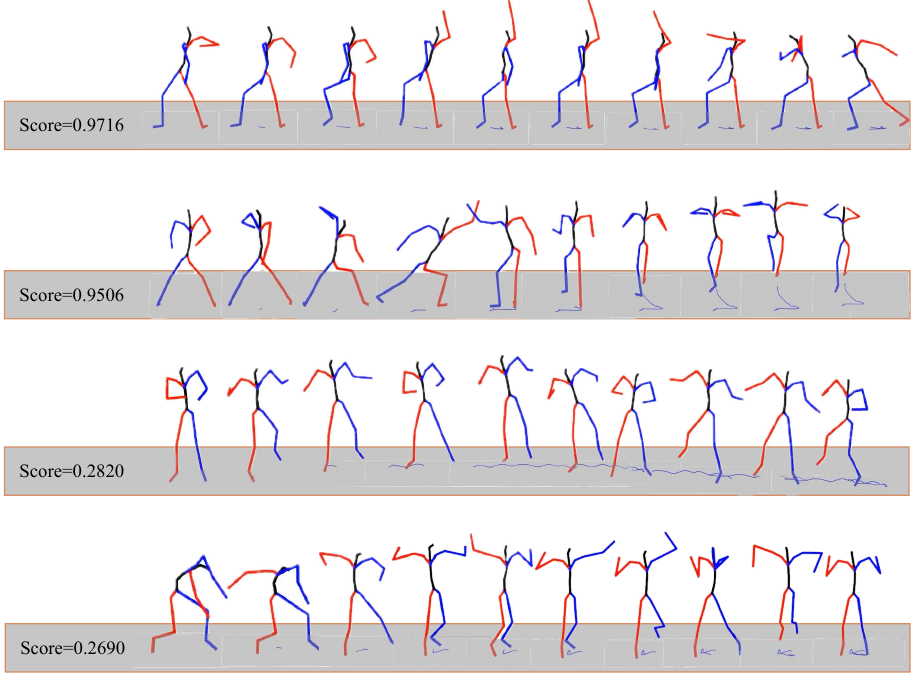}
	\caption{\textbf{Qualitative results of Music-Motion Retrieval.} For the \textit{Pop} music genre, higher similarity scores indicate greater correspondence between the retrieved dance motions and the input music.}
        \label{fig:mmr_results}
\end{figure*}

\noindent\textbf{Experiments.}  
We conducted both qualitative and quantitative experiments to evaluate the performance of the MMR module. As shown in Table~\ref{tab:mmr_exp}, MMR demonstrates exceptional retrieval capabilities. Visualized retrieval results further validate this observation. For comparison, we provide two music samples, each paired with two high-similarity dance sequences and two low-similarity dance sequences. Figure~\ref{fig:mmr_results} and the demo examples illustrate that our retrieval results align better with the beat and rhythm. In these examples, higher similarity scores indicate a stronger correlation between the music and the retrieved dance motions. 

\begin{table}[h]
    \resizebox{\linewidth}{!}{
        \begin{tabular}{lcccccc}
            \toprule [1pt]
             Retrieval Task
            & R@1 $\uparrow$ & R@2 $\uparrow$  & R@3 $\uparrow$  & R@5 $\uparrow$ & R@10 $\uparrow$ & MedR $\downarrow $  \\
            \noalign{\smallskip}\midrule[0.5pt]
            \noalign{\smallskip}
            Music-Motion Retrieval & 42.04 & 56.95 & 64.93  & 73.94 & 84.62 & 2.00\\
            Motion-Music Retrieval & 42.00 & 57.52 & 65.62  & 74.48 &84.22 & 2.00\\
            \bottomrule [1pt] 
        \end{tabular}
        }
        \caption{\textbf{Retrieval results on the dance dataset.} Both Music-Motion Retrieval and Motion-Music Retrieval tasks maintain Recall@1 performance above 40\%.}
    \label{tab:mmr_exp} 
    \vspace{-1em}
\end{table}

\noindent\textbf{Supervised MAGM.} The pretrained MMR module provides music-motion alignment supervision for the MAGM dance generation pipeline. A critical challenge arises from the discrete token inputs to MAGM, while the MMR's alignment losses $\mathcal{L}_{\text{Align-body}}$ and $\mathcal{L}_{\text{Align-whole}}$ require continuous motion representations for contrastive learning. How can we bridge this gap and enable gradient propagation through the non-differentiable process? As illustrated in Fig.~\ref{fig:gpt}, we address this issue in three steps. First, the discrete discrete tokens are decoded into continuous motion sequences $M \in \mathbb{R}^{T \times D_m}$ via the hierarchical residual vector quantization decoder $\mathcal{D}_{\text{whole}}$; second, the reconstructed motion $M$ is encoded through the MMR's motion encoder $\mathcal{E}_{\text{motion}}$ to obtain latent code $\mathbf{z}$, enabling the computation of InfoNCE loss with music features $\mathbf{c}$; and third, to enable end-to-end training despite discrete token sampling, we employ Gumbel-Softmax relaxation~\cite{jang2016categorical} during token generation, which provides a continuous approximation of the discrete sampling process and allows gradient flow through the otherwise non-differentiable quantization step, with the temperature parameter $\tau$ annealed during training to progressively sharpen the distribution. 

\section{User Study Details}
\label{supl:user_study_details}

A/B videos are randomly sampled clips from different datasets or generated by different methods, presented to users for comparison and evaluation. As shown in Figure~\ref{fig:user_study}, after watching dance videos A and B, participants were asked to answer the following questions:
\begin{itemize}
    \item Please rate A/B based on your level of preference. 
\end{itemize}
\begin{itemize}
    \item Considering only the body movements of A/B, please rate based on your level of preference. 
\end{itemize}
\begin{itemize}
    \item Considering only the hand movements of A/B, please rate based on your level of preference. 
\end{itemize}
\begin{itemize}
    \item Considering only facial expressions, how well does A/B convey the emotional tone of the music? Please rate.
\end{itemize}
\begin{itemize}
    \item How well does A/B align with the rhythm of the music? Please rate.
\end{itemize}

We conducted a user study on the dance datasets, selecting four different music genres from the \textit{SoulDance} dataset. For each genre, a random music-dance sequence was chosen and compared with sequences of the same genre from the AIST++~\cite{li2021ai} and FineDance~\cite{li2023finedance} datasets. Participants were then asked to rate various performance aspects for each comparison.

In addition, we performed a user study on different dance generation methods. Under identical music conditions, we conducted pairwise comparisons between results generated by \textit{SoulNet} and those produced by FACT~\cite{li2021ai}, Bailando~\cite{siyao2022bailando}, EDGE~\cite{tseng2023edge},  FineNet~\cite{li2023finedance} and ground truth. Participants rated different aspects of each generated dance sequence. Training and generation were carried out separately on both the AIST++~\cite{li2021ai} and SoulDance datasets.



\begin{figure}[h]
	\centering
	\includegraphics[width=0.6\linewidth]{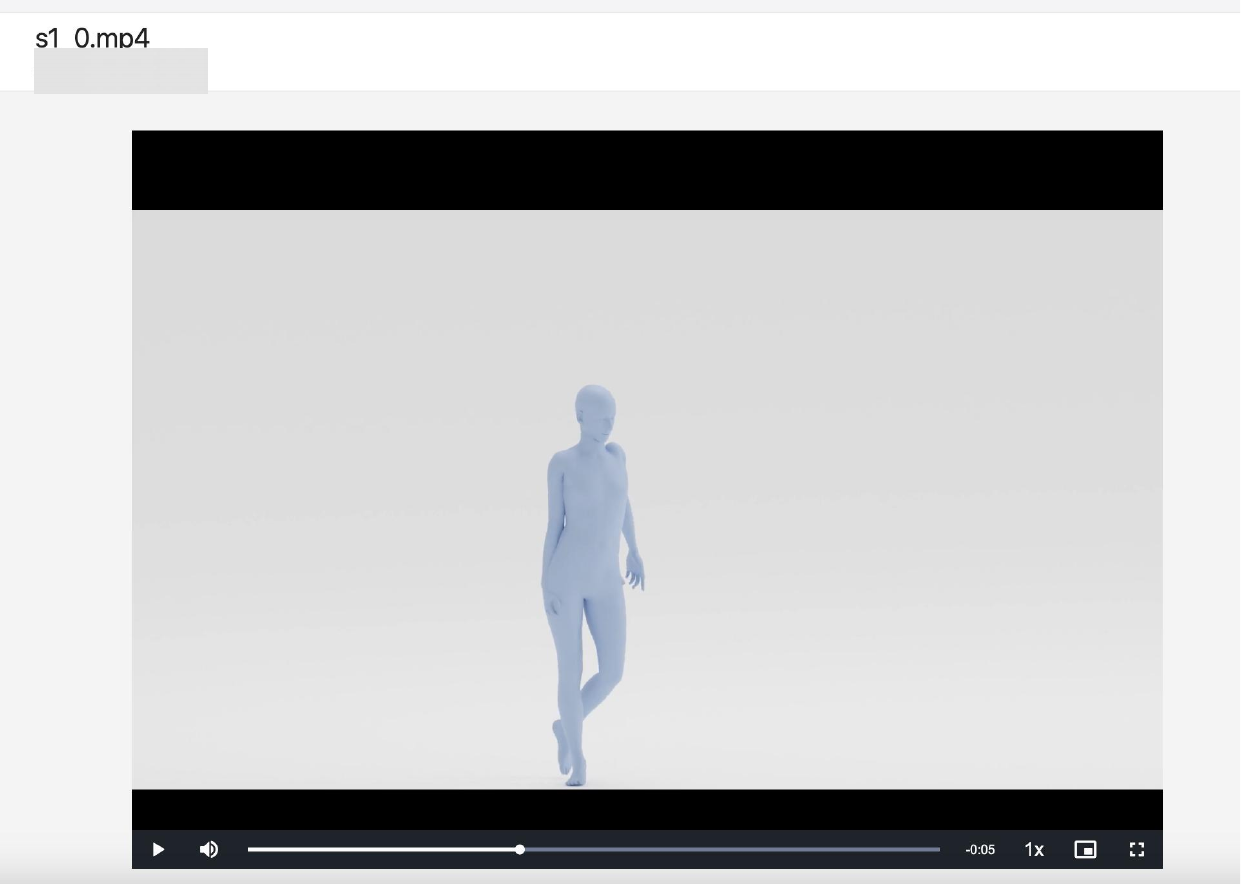}
	\caption{\textbf{Screenshot of video page in the user study.} The interface provides independent A/B video links, allowing users to view each corresponding video separately.}
        \label{fig:user_study_video}
\end{figure}

\begin{figure}[th]
	\centering
	\includegraphics[width=0.5\linewidth]{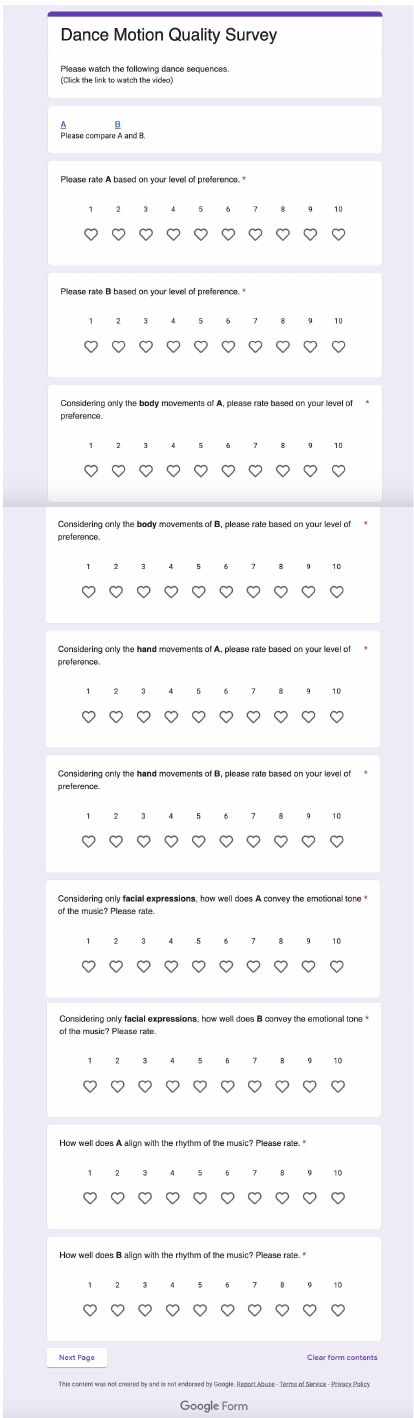}
	\caption{\textbf{User interface of our surveys.} The interface presents a set of questions alongside two videos, A and B. Screenshots of the videos linked in the survey are shown in Figure~\ref{fig:user_study_video}.}
        \label{fig:user_study}
\end{figure}

\end{document}